\baselineskip=12pt

\def\adots{\mathinner{\mskip1mu\raise1pt\hbox{.}\mskip2mu\raise4pt\hbox{.}\mskip2mu\raise7pt\vbox{\kern7pt\hbox{.}}\mskip1mu}}
\def\diag{\mathop{\rm diag}\nolimits}

\mathchardef\bfplus="062B
\mathchardef\bfminus="067B
\font\title=cmbx10 scaled\magstep5
\font\chapter=cmbx10 scaled\magstep4
\font\section=cmbx10 scaled\magstep2
\font\subsec=cmbx10 scaled\magstep1

\def\~#1{{\accent"7E #1}}
\def\sqr#1#2{{\vcenter{\hrule height.#2pt \hbox{\vrule width.#2pt height#1pt \kern#1pt \vrule width.#2pt}\hrule height.#2pt}}}

\def\hk#1#2{{\vcenter{\hrule height0.0pt \hbox{\vrule width0.0pt \kern#1pt \vrule width.#2pt height#1pt}\hrule height.#2pt}}}

\centerline{\chapter Modelling Sex Ratio and Numbers for} 
\centerline{ \chapter Translocation in Meta-Population Management}
\vskip 24pt
\noindent Peter R. Law\hfil\break
1 Mack Place\hfil\break
Monroe, NY 10950\hfil\break
U.S.A.\hfil\break
email: prldb@member.ams.org\hfil\break
\vskip 24pt
\noindent {\bf ABSTRACT}\hfil\break
The management of endangered species as metapopulations is becoming increasingly common. Diverse aspects of metapopulation dynamics and management have received attention in recent years. In particular, translocation of individuals between subpopulations of a metapopulation is practiced, or envisaged, for a variety of reasons and requires careful consideration. Linklater (2003) proposed that the number of individuals of each sex translocated into a target population for the purposes of maintaining genetic diversity could be chosen on the basis of parental investment theory. In this paper, following basic ideas in the parental investment literature, I propose a simple model to capture Linklater's proposal and provide a thorough mathematical analysis of the model. Granted the necessary species-specific biological information which would determine the model parameters in any instance of application, the analysis indicates that a practical algorithm can be constructed to generate the model's predictions for the optimal translocation.
\vfill\eject
\noindent{\section 1. INTRODUCTION}
\vskip 12pt
Many endangered species now exist in fragmented habitat and are managed as meta-populations, the subspecies of black rhino {\it Diceros bicornis} of Africa providing notable examples. This situation poses many problems as regards reserve design, population dynamics, and management. In particular, reduced possibility, or even impossibility, of natural dispersal in such circumstances reduces genetic flow between subpopulations and may decrease productivity and increase risk of local extinction due to reduced fitness and stochastic effects (e.g., Soul\'e 1987). 

Translocation of individuals is one option for maintaining genetic flow between artificially isolated subpopulations of a given (sub)species. Given the inherent risks and practical challenges (e.g., Maguire et al. 1987, Brett 1998, Miller et al. 1999, Fischer \& Lindenmayer 2000) involved in this practice, choosing strategies that optimize the reproductive return from translocation efforts and enhance metapopulation persistence is sound policy. A variety of issues have received theoretical treatment (the following list and literature are not intended to be exhaustive): which life stages are optimal for translocation (Hearne \& Swart 1991, Robert et al. 2004); the role of captive populations (Tenhumberg et al. 2004, McPhee \& Silverman 2004); decision making in the face of limited resources and uncertainties (Maguire 1986, Haight et al. 2000); metapopulation dynamics modelling (Hearne \& Swart 1991, Wootton \& Bell 1992, Lubow 1996); rate of genetic flow required (Wang 2004). 

As some of these studies emphasize, the goal of translocation is not just to merely supplement numbers but also to enhance reproductive performance and maintain genetic flow. Hence, the reproductive performance of the translocated individuals in the target population should be a primary concern and a measure of the success of a translocation. Linklater (2003) argued that managers planning a translocation can be viewed as parents, constrained by limited resources, investing in offspring and suggested that translocations be modelled in terms of parental investment theory. Application to a specific species will then depend on assumptions regarding reproductive behaviour and how individuals respond to investment.

In a polygynous, sexually reproducing species, all females are expected to reproduce while only the fittest males are likely to obtain reproductive opportunities. Trivers and Willard (1973) argued that for such species natural selection should result in a mechanism that enables a mother to influence, presumably at conception, the sex of her offspring so as to gain the optimal return on her investment in that offspring from the resources available to her at that time.  Specifically, they predicted that when a mother is in poor condition she should produce daughters rather than sons and invest more in the daughters she does produce, while when in good condition she should produce sons rather than daughters and invest more in the sons she produces. These predictions presume that the condition of the mother determines her ability to invest in her offspring, that investment in offspring translates into the condition of that offspring at sexual maturity, and reflect that, in the polygynous context of the considerations, males require a competitive edge to gain mating opportunities so only a mother in good condition is in a position to produce such a son while mothers in poor condition are better off investing their more limited resources in females.

While there has been some controversy about what Trivers and Willard actually meant, see for example Carranza (2002), Cameron and Linklater (2002) have argued forcefully for the interpretation outlined above. Moreover, Cameron et al. (1999) and Cameron \& Linklater (2000) found evidence for the Trivers-Willard (TW) effect in wild horses in New Zealand consistent with their interpretation. Although there have been many studies of the TW effect, with mixed results, Cameron (2004) conducted a meta-analysis of mammalian sex-ratio studies and reported that much of the inconsistency in results can be attributed to the variety of indices of maternal condition employed. Studies that utilized indices of maternal condition near conception provided almost unanimous support for the TW effect. In addition, Cameron (2004) described a possible mechanism for facultative adjustment of sex ratio, through the effect of blood-glucose levels on blastocysts, which would be consistent with the TW effect. Subtleties of life history of a species may confound the prediction of which sex benefits more from maternal investment (Clark 1978, Leimar 1996, Sheldon \& West 2004) but does not necessarily undermine the general argument of Trivers \& Willard.

Linklater (2003) specifically proposed that the TW effect could be adapted as a model for how managers should invest resources in translocation operations for meta-population management of polygynous species such as black rhino; specifically, that the number of males and females to be released into a given target population, and the partition of resources amongst them, should be modelled as a function of the resources available using the TW effect as the underlying principle. Optimization of such a model would then provide the recommended translocation strategy for managers to adopt in a given circumstance. The approach envisaged by Linklater would constitute an important component of metapopulation management.

In this paper, I present a simple model of reproductive returns on investment in translocations. Linklater's specific proposal would be incorporated by appropriate choice of return functions for males and females. I describe the model in \S\S 2--4, basic elements of which are adapted from ideas in Frank (1987). Solution of the model involves two optimization problems, which are solved in \S\S 5--9. These solutions, along with numerical case studies, indicate both the generality and practicality of the model. 
\vskip 24pt
\noindent {\section 2. INGREDIENTS OF THE MODEL}
\vskip 12pt
The aim is to model the return on investment in a translocation, where, by {\sl return\/}, I mean a measure of the reproductive success of the translocated animals (which reflects the transfer of alleles into the population). 

By {\sl fixed costs\/} are meant necessary costs independent of the number of animals translocated which do not yield any return. In the modelling, the available resources shall mean the resources after the exclusion of fixed costs. 

{\sl Packaging costs\/} are, by definition, costs per individual released which do not contribute to the animal's success, i.e., there is no return on packaging costs, but these costs must be included in the model. An example might be transportation costs, but in general packaging costs will be specific to each translocation effort and may well be gender specific.

I assume that the resources invested in an individual yield a reproductive return that can be modelled by a suitable {\sl return function\/} of said resources. I further assume that the return function is fixed for each gender in any given application. In reality, for a given species, a particular life stage may be optimal for translocation. One might therefore always choose individuals from that life stage or, with greater difficulty, allow different return functions for different life stages of each gender in the modelling.

Let $x$ denote the proportion of available resources to be invested in males and $z$ the proportion invested in females, whence $x+z=1$. I denote the return function for a male by $m(x)$ and that for a female by $f(z)$. The choice of return function for each gender models expected response to investment of resources and behavioural assumptions about the species; in particular, appropriate choices incorporate the TW effect into the model. Beyond the choice of return functions, there are three further steps in the modelling process: determining how to invest a given level of resources in a given gender; combining the results of the previous step to find the overall return from investment in both genders; computing the optimal overall return and hence the recommended translocation strategy.

Note that even if there were an ideal return function that models the behaviour of individuals of a given gender of a given species in the context of unlimited resources, the return function employed in a given application must reflect the available resources. In particular, while returns no doubt level off in the face of unlimited resources, in the context of limited resources such behaviour may not be evident. This fact may mean that functions which are unrealistic return functions in the context of unlimited resources might serve over limited ranges of resources. It may also be the case that one knows only an approximation to the actual return function.

For this reason, and for some mathematical generality, I treat a collection of candidate return functions which are nondecreasing and simply characterized by their rates of change, specifically, by their first and second derivatives. Each be regarded as approximations to the sigmoid in appropriate circumstances.
\vskip 24pt
\noindent {\section 3. INVESTING IN A SINGLE GENDER}
\vskip 12pt
In this section I will employ the notation appropriate for males; it is modified for females in a straightforward manner. When resources $x$ are partitioned into $x = \sum_{i=1}^n\,x_i$, with $x_i$ applied to the $i$'th of $n$ individuals, the total return is 
$$M(x;n;x_1,\ldots,x_n) := \sum_{i=1}^n\,m(x_i).\eqno(3.1)$$
For given $x$, also define
$${\cal M}(x) := \max_{\scriptstyle (n;x_1,\ldots,x_n) \atop \scriptstyle x_1+\cdots+x_n=x}M(x;n;x_1,\ldots,x_n).\eqno(3.2)$$
${\cal M}(x)$ is the return from the optimal assignment of resources $x$ to males with individual return function $m(x)$. Similarly, one defines a function ${\cal F}(z)$ for females.
\vskip 12pt
\noindent {\subsec 3.a Step-Function Returns}
\vskip 12pt
Suppose each male requires a certain minimal investment $a$, beyond the packaging cost $d$, before any return on investment occurs and that any investment at and beyond $a+d$ results in a fixed return $r_m$. Then the return function for a single male is a step function:
$$m(x) =\cases{r_m,& if $x \in [a+d,1]$;\cr 0,& if $x \in [0,a+d)$.\cr}\eqno(3.{\rm a}.1)$$
By the nature of this individual return function, for resources $x$ the optimal strategy is to release $n_m$ males where $n_m=[x/(a+d)]_i$, with $[\ ]_i$ denoting the integral part of the bracketed number. It follows that
$${\cal M}(x) = r_m\left[{x \over a+d}\right]_i.\eqno(3.{\rm a}.2)$$
\vskip 12pt
\noindent {\subsec 3.b Linear Returns}
\vskip 12pt
Suppose the returns are linear functions of investment beyond the packaging costs:
$$m(x) = \cases{a(x-d)& if $x \in [d,1]$;\cr 0& if $x \in [0,d]$.\cr}\eqno(3.{\rm b}.1)$$
If resources $x$ are partitioned amongst $n$ males, then
$$M(x;n;x_1,\ldots,x_n) = \sum_{i=1}^n\,m(x_i) = \sum_{i=1}^n\,a(x_i-d) = a(x-nd).\eqno(3.{\rm b}.2)$$
Due to the nonzero packaging costs, the optimal strategy is to choose $n=1$, whence
$${\cal M}(x) = m(x).\eqno(3.b.3)$$
\vskip 12pt
\noindent{\subsec 3.c Increasing-Marginal-Return (IMR) Functions}
\vskip 12pt
In this case the return function is increasing with increasing marginal returns (the graph is concave up), which may be characterized as follows: the function is zero on $[0,d]$ and on $[d,1]$ satisfies
$$m(d) = 0 \hskip .75in m'(x) > 0 \hskip .75in m''(x) > 0.\eqno(3.{\rm c}.1)$$

One can use the result of the Appendix A to solve the optimization problem. As usual, it makes sense to assign each male invested in at least $d$ of the resources. Suppose one has so invested in $n$ males. Put $y_j = x_j -d > 0$, $j=1,\ldots,n$. Since $m'$ is injective ($m'' > 0$ implies $m'$ strictly increasing) on the relevant domain, (A.5) has a unique solution $y_1 = \cdots = y_n$. But, as $m'' > 0$, the Hessian in (A.6) is positive definite at this critical point, i.e., one has a local minimum, not a maximum. So equal apportionment is the least optimal strategy in this case when resources are divided amongst several individuals. It follows from Appendix A that the optimal strategy is to invest in a single individual, whence ${\cal M}(x)=m(x)$. The linear return of 3.b is a limiting instance of this case.
\vskip 12pt
\noindent {\subsec 3.d Diminishing-Marginal-Return (DMR) Functions}
\vskip 12pt
In this case the return function is increasing with decreasing marginal returns (the graph is concave down), characterized by a return function $m$ which is zero on $[0,d]$, while on $[d,1]$, 
$$m(d) = 0 \hskip .75in m'(x) > 0 \hskip .75in m''(x) < 0.\eqno(3.{\rm d}.1)$$
One may require the function to have a horizontal asymptote, in which case $\lim_{x \to \infty}m'(x) = 0$. 

From Appendix A, since $m'' < 0$ on $(d,1]$, $m'$ is injective on this subinterval. Hence, for $n$ satisfying (A.2), (A.5) again has a unique solution: $y_1=\cdots=y_n$ is the only critical point and the Hessian (A.6) is negative definite, i.e., this critical point is a local maximum. 

Thus, equal apportionment of resources $x > d$ is the optimal strategy amongst $n$ individuals, for $n$  satisfying (A.2). For fixed $x$ and $n$, the optimal return is therefore
$$M(x;n) := nm(x/n).\eqno(3.{\rm d}.2)$$
It remains to optimize with respect to $n$ subject to (A.2):
$${\cal M}(x) = \max_{0 < n < x/d}M(x;n)\eqno(3.{\rm d}.3)$$
with the understanding that when $x \leq d$ there are no such $n$ and ${\cal M}(x) = 0$, i.e., ${\cal M}(x) = 0$ on $[0,d]$. Note that on $(d,2d]$, $n=1$ is the only possibility, so in fact
$${\cal M}(x) = m(x) \hbox{ on } [0,2d].\eqno(3.{\rm d}.4)$$

Consider the function $\phi_x(v) := vm(x/v)$ for fixed $x \in (d,1]$. The restriction $vd < x$ entails one may take the domain of $\phi_x$ to be $(0,x/d)$. With $v \in (0,x/d)$, $x/v \in (d,\infty)$, whence $\phi_x$ is smooth on its domain if $m$ is smooth on $(d,\infty)$, which is presumed. Differentiating with respect to $v$ yields:
$$\displaylines{\phi_x'(v) = m(x/v) - \left({x \over v}\right)m'(x/v);\cr
\hfill\llap(3.{\rm d}.5)\cr
\phi_x''(v) = {x^2 \over v^3}m''(x/v).\cr}$$
Since $m'' < 0$ on $(d,\infty)$, and $x$ and $v$ are positive, then $\phi_x'' < 0$ on its domain, i.e., $\phi_x$ is concave down on $(0,x/d)$, whence has a local maximum which is its global maximum. Let $v_x$ denote the value of $v$ for which this maximum occurs, i.e., $v_x$ solves $\phi_x'(v_x) = 0$. Putting $t_x := x/v_x \in (d,\infty)$, then
$$m(t_x) = t_xm'(t_x);\qquad\hbox{equivalently}\qquad m'(t_x) = {m(t_x) \over t_x}.\eqno(3.{\rm d}.6)$$
This equation has a simple geometric interpretation; namely, the tangent line to $m$ at $t_x$ coincides with the line through the point $\bigl(t_x,m(t_x)\bigr)$ and the origin (since they have a point in common and the same slope). As $m$ is concave down for $x > d > 0$ and zero on $[0,d]$, there is a unique line through the origin and tangent to $m$. This geometric fact characterizes $t_x$ and indicates it is independent of $x$. Thus, one may write $t$, defined by
$$m'(t) = {m(t) \over t}.\eqno(3.{\rm d}.7)$$

If one did not require integral values, then the optimal $n$ for given $x$ would be $n_x = v_x = x/t$, which is a simple linear function of $x$. Since, however, $n$ must be integral, because of the concave down shape of $\phi_x$, the optimal $n$ for given $x$ is the integer either side of $x/t$ yielding the largest return:
$$n_x = \left[{x \over t}\right]_i\hbox{ or } \left[{x \over t}\right]_i+1\qquad\hbox{according as}\qquad \left[{x \over t}\right]_i m\left({x \over [x/t]_i}\right) > \left(\left[{x \over t}\right]_i+1\right)m\left({x\over [x/t]_i+1}\right)\hbox{ or not}.\eqno(3.{\rm d}.8)$$
Thus, for a given $x$ there are just two candidates for $n_x$. Notice that for $d < x < t$, $[x/t]_i = 0$, whence $n_x$ must be 1. In particular, if $t \geq 1$, then $n_x=1$ for all $x \in (d,1]$.
\vskip 12pt
\noindent {\bf 3.d.9 Observation}\hfil\break
Given $m$ as in (3.d.1), if the solution $t$ of (3.d.7) satisfies $t \geq 1$ then
$${\cal M}(x) = m(x).$$ 
\vskip 12pt
If $t < 1$, however, as $x$ increases, so does $x/t$ and so the {\sl trend\/} in $n_x$ should be increasing. Note that although $t$ is independent of $x$, $\phi_x$ is not; in particular, its critical point $v_x = x/t$ changes with $x$, as does the domain $(0,x/d)$. In general, ${\cal M}(x) = n_xm(x/n_x)$.

Explicit examples of (3.d.1) are: on $[d,1]$:
$$h(x) = K(x-d)^s\qquad K > 0,\qquad 0 < s <1 \hskip 1.25in g(x) = K\left(1 - \left({d \over x}\right)^s\right)\qquad s > 0.\eqno(3.{\rm d}.10)$$
The functions $h$ are derived from the cumulative distribution function for certain $\beta$-distributions and were considered by Frank (1987). They increase without bound. The functions $g$ exhibit a horizontal asymptote at $y=K$; consequently, $g$ is bounded. Note also that for $s \gg 1$, $g(1) \approx K$, so by an appropriate choice of $s$ one can ensure that $g$ approaches its bound on the interval $[0,1]$.

For the function $h$, the solution of (3.d.7) is $t=d/(1-s) > d$ but increases without bound as $s\rightarrow 1$. In particular, $t \geq 1\ \Leftrightarrow\ s \geq 1-d.$ For such $s$, (3.d.9) is valid. Note that as $s \to 1$, the function $h$ is almost linear, whence the result just noted is consistent with the results for linear return functions. In general, $v_x = (1-s)(x/d)$.

For the function $g$, the solution of (3.d.7) is $t = \root s \of{s+1}\,d > d$. Taking the logarithm and using L'H\^opital's rule shows that $\lim_{s \to \infty} \root s \of{s+1} = 1$ while $\lim_{s \to 0} \root s \of{s+1} = e$, i.e., $ d < t < de$ as $s$ ranges from $\infty$ down towards 0. If $\root r \of{r+1}\,d = 1$, for some $r \in (0,\infty)$, then $\root s \of{s+1}\,d \geq 1$ for $0 < s \leq r$. But $t \geq 1$ requires $1 < de$, i.e., $d > 1/e \approx 0.37$. In particular, if $d \leq 1/e$, then $t < 1$. So, if $d$ is less than 0.37, then (3.d.9) is inapplicable. In general, $v_x = [1/\root s \of{s+1}](x/d)$.
\vskip 12pt
\noindent {\subsec 3.e Sigmoid Return Functions}
\vskip 12pt
One need not restrict to the logistic $g(x) = 1/[1+\exp(-x)]$. One only requires, on $[d,1]$,
$$\displaylines{m(d) = 0 \hskip .75in m'(x) > 0\cr
\hfill\llap(3.{\rm e}.1)\cr
\hbox{and for some $p$, $d < p <1$,}\qquad m''(x) \cases{> 0,& for $x \in [d,p)$;\cr < 0,& for $x \in (p,1]$.\cr}\cr}$$
Demanding a horizontal asymptote requires $\lim_{x \to \infty}m'(x) = 0$. Assuming that $m$ is $C^2$, then 
$$m''(p) = 0.\eqno(3.{\rm e}.2)$$

For any given $x$, and $n$ satisfying (A.2), critical points of $M(x;n,x_1,\ldots,x_n)$ satisfy (A.5). If $x_1,\ldots,x_n$ are all less than $p$, then as $m'$ is injective on $(d,p)$, as in (3.3) $x_1=\cdots=x_n$ and such a critical point gives a local minimum of the return function. This phenomenon obviously occurs for $x < p$ for which $m$ is exactly as in (3.e). Assuming continuity of ${\cal M}$, for $x \leq p$, ${\cal M}(x) = m(x)$. 

When $x > p$, it becomes possible to have solutions of (A.5) with $x_j \not= x_k$, in which case $x_j < p$ and $x_k > p$ say. But the Hessian (A.6) of such a critical point is indefinite. By (A.5), $m'(x_j) = m'(x_k)$. To first order, diverting a small amount of the resources $x_k$ from the $k$'th individual to the $j$'th individual exactly balances. But, to second order, since $m''(x_k) < 0$ while $m''(x_j) > 0$, the return from individual $k$ will decline while that from individual $j$ will increase. The overall change in return will depend, to second order, on the relative magnitudes of $m''(x_k)$ and $m''(x_j)$. Such critical points are saddle points. 

If $x$ is large enough that $x/n > p$, then $x/n = x_1=\cdots=x_n$ is a solution of (A.5) and (A.6) indicates it yields a local maximum amongst all partitions of $x$ into $n$ amounts (including those partitions for which some $x_j$ are less than $p$). Unlike in (3.d), a local maximum may fail here to be the absolute maximum because of the partitions containing $x_j < p$ (the saddle-point critical points provide stationary points where the function may start to increase in certain directions and thereby ultimately exceed the value of the function at the local maximum). If so, the maximum for this value of $n$ must occur on the boundary of the domain and the argument of Appendix A still entails that the optimal strategy for investing resources $x$ is equal apportionment amongst $n$ individuals for some $n$ satisfying (A.2).

In summary, the optimal strategy when $x > p$ is amongst the strategies $nm(x/n)$, $nd < x$, $x/n > p$. Note that when $x < p$, there is no such $n$, whence $n=1$ is the optimal strategy, so the earlier deduction for the case $x < p$ may be included in this characterization. Hence, the situation may be regarded as a generalization of that in (3.d).

Indeed, as in (3.d), one can consider the function $\phi_x(v) := vm(x/v)$. The calculations (3.d.5) are unchanged. Again, one finds that there exists a $t$ satisfying (3.d.7) which yields the unique critical point for $\phi_x$. Put $v_x := x/t$ as before. In the present circumstances,
$$\phi_x''(v_x) =  {x^2 \over v_x^3}m''(t).$$
By the shape of the sigmoid graph, $t > p$, whence $m''(t) < 0$. Thus, $v_x$ is the unique local maximum of $\phi_x$. Moreover, since only those $n$ for which $x/n > p$ are of interest here, one can restrict $v$ so that $x/v > p$, i.e., $\phi_x$ is indeed concave down on the domain of interest. So, as in (3.d), one can solve (3.d.7) for $t$ and then the optimal strategy is as in (3.d.8). In particular, $x \leq t\ \Rightarrow n_x = 1$ and (3.d.9) is valid for sigmoid returns too.

Observe that if $p$ is close to 1, then the sigmoid function is approximately as in (3.3), while if $p$ is close to 0, then the sigmoid is approximately as in (3.d). Moreover, the step-function return may be viewed as an extreme limiting form of the sigmoid, while the linear return is a limiting form of both (3.c) and (3.d) and thus an approximation to certain sigmoids. One would therefore expect some unifying perspective on the results obtained in this section. For example, the optimal strategy for step functions was similar to that in (3.d) being also equal apportionment but with the difference that it was amongst the maximum possible number of individuals. That said, the line through the origin that touches the nonzero part of the step function goes through the point $x=a+d$, i.e., for the step function, although (3.d.7) is not applicable, the geometric construction of $t$ would give $t=a+d$, which is consistent with (3.d.7) in that $t$ is the amount of resources one would ideally equally apportion (without the integral requirement imposed on $v$) amongst individuals. 

Indeed, a unifying viewpoint is as follows. For any return function m(x), the quantity m(x)/x is the return per unit resource. This quantity is maximal for the point of the graph of m(x) where the steepest line through the origin touches the graph. Denoting by $t$ this value of $x$, $m(t)/t$ is the maximal value of the return per unit resource. Since optimal investment in the cases studied here never involves unequal partition of resources amongst several individuals, one may restrict attention to equal division of resources $x$ into amounts $y$ amongst individuals, which yields a return of $(x/y).m(y) = x.m(y)/y$, and the optimal strategy would be ideally $xm(t)/t$. Since the number of individuals must be integral, the actual optimal strategy is the best approximation to this ideal strategy. This argument reproduces, at least, the final result in each case (a)--(e) above. For step-function returns, $t = a+d$; for IMR functions (including linear), $t = 1$ and, although equal division of resources amongst several individuals is not actually optimal in any sense, the optimal strategy is to invest all resources $x$ in a single individual, which does amount to $t=1$; for DMR and sigmoid functions, $t$ is characterized by (3.d.7) (when $m(x)$ is smooth, elementary calculus shows that $m(x)/x$ has a critical point satisfying (3.d.7) which is a local maximum when $m'' < 0$ at the critical point).

It is noteworthy that when one computes ${\cal M}(x) = n_xm(x/n_x)$ for (3.d.10) and the example sigmoid function of \S 6.c, the resulting graph of ${\cal M}(x)$ appears to represent a continuous function of $x$. I consider this issue in Appendix B, which results, for some return functions, in a simple procedure for obtaining $\cal M$.
\vskip 24pt
\noindent {\section 4. THE MODEL}
\vskip 12pt
In order for male and female reproductive returns to be comparable, male returns must be weighted by reproductive opportunities, which is achieved by multiplying by the appropriate sex ratio of the target population. Let $S := \phi/\mu$ denote the ratio of reproductively active females to reproductively active males in the target population prior to the release and let $S_P$ denote the same ratio after the release, where the subscript indicates that $S_P$ depends on the actual partitioning of resources through the numbers of each gender released. While the interpretation of $\phi$ and $\mu$ may depend on the species considered, since the aim of any translocation is to introduce breeding stock into the target population, I will assume that each translocated animal is included in $S_P$.

The model I pose is as follows. Let $d$ denote the packaging costs for males and $\delta$ that for females. With notation as in (3.1), define
$$R_1(x) := \max_{\scriptstyle x_1+\cdots+x_n=x \atop \scriptstyle z_1+\cdots+z_k=1-x}[M(x;n;x_1,\ldots,x_n)S_P + F(1-x;k;z_1,\ldots,z_k)],\eqno(4.1)$$
where the maximization is over all permissible partitions: 
$$\displaylines{n < {x \over d},\hskip .75in  \sum_{i=1}^n\,x_i =x,\hskip .75in y_i := x_i-d \in (0,x-nd);\cr
k < {1-x \over\delta},\hskip .75in \sum_{j=1}^k\,z_j=1-x,\hskip .75in w_j := z_j-\delta \in (0,1-x-k\delta).\cr}$$
Investment in a individual which yields no return is suboptimal for that gender. Moreover, such individuals should not count in $S_P$ and so cannot enhance overall returns. Hence, such investment is suboptimal across genders and therefore need not be considered.

For target populations with a large number of both males and females relative to the number of releases, $S_P \approx S$ and a simplified model results by replacing $S_P$ by $S$. In this case, since $S$ is constant, the optimization in (4.1) splits into separate optimizations of each summand, each of which then assumes the form as in (3.2) and (4.1) becomes
$$R_2(x) := {\cal M}(x)S + {\cal F}(1-x).\eqno(4.2)$$
One might wonder whether (4.1) ever reduces to
$$R_3(x) := {\cal M}(x)S_x + {\cal F}(1-x),\eqno(4.3)$$
where $S_x$ denotes the sex ratio of the post-release population for the optimal divisions of resources $x$ amongst males and of $1-x$ amongst females. I shall consider this question case by case. I shall refer to (4.1) as Model 1, (4.2) as Model 2, and (4.3) as Model 3.

It follows from the considerations in \S 3 that for each $x \in [0,1]$ there are only finitely many possible strategies over which one must optimize to form $\cal M$, $\cal F$, $R_2$ and $R_1$, whence each of these functions is well defined. When $m$ and $f$ are continuous, it follows, as argued in Appendix B, that $\cal M$ and $\cal F$ are each continuous, whence $R_2$ is continuous. Thus, Model 2 is well defined in that $R_2$ always possesses a maximal return on $[0,1]$. The continuity of $R_1$ depends, however, on the factor $S_P$, which may vary discontinuously. Thus, $R_1$, while bounded on $[0,1]$, if discontinuous may not achieve its supremum (least upper bound). I will discuss this issue in the context in which it arises.

In the case of (4.2), if ${\cal M}(x)$ and ${\cal F}(1-x)$ are smooth functions of $x$, then $R_2'(x) = {\cal M}'(x)S - {\cal F}'(1-x)$ and the critical points are given by
$${\cal M}'(x)S = {\cal F}'(1-x),\eqno(4.4)$$
with a local maximum ensured by
$$0 > R_2''(x) = {\cal M}''(x)S + {\cal F}''(x).\eqno(4.5)$$
The optimal strategy would then be at such a local maximum or at $x=0$ or $x=1$. Since (4.4--5) require analytic expressions for $\cal M$ and $\cal F$, they are rarely a practical method of solution.

In the following sections, I conduct case studies of Models 1 and 2. The purpose of these studies is to explore the mathematical behaviour of the model and present solutions to various scenarios. The biological relevance of these case studies is not explicitly considered, but see the discussion. For notational convenience, when (fe)males, are modelled with a return function as in 3a -- e, I shall refer to step-function, linear, IMR, DMR, or sigmoid (fe)males, respectively. I do not explicitly study return functions of type 3c in the following case studies though it would not be difficult to do so.  The results for linear returns indicate what to expect for returns of type 3c.
\vskip 24pt
\noindent {\section 5. STEP-FUNCTION MALES AND FEMALES}
\vskip 12pt
The individual return function for males is as in (3.a.1), while the individual return function for females is:
$$f(z) =\cases{r_f,& if $x \in [\alpha+\delta,1]$;\cr 0,& if $x \in [0,\alpha+\delta)$.\cr}\eqno(5.1)$$
Such return functions might be appropriate for modelling in situations where there is a regime of fixed investment per individual and only a knowledge of average returns modelled as a flat return; in particular, translocations called hard releases.

As we saw in (3.a), the resulting optimal strategy for assigning resources $x$ to males is
$${\cal M}(x) = r_m\left[{x \over a+d}\right]_i.\eqno(5.2)$$
Similarly, the optimal strategy for assigning resources $z$ to females is 
$${\cal F}(z) = r_f\left[{z \over \alpha + \delta}\right]_i.\eqno(5.3)$$
Observe that if $a+d+\alpha+\delta > 1$, then the resources are sufficient only for investment in either males or females. If, further, both $a+d \leq 1$ and $\alpha+\delta \leq 1$, then both options are available. For an investment exclusively in females the optimal return is $r_f.[1/(\alpha+\delta)]_i$. For an investment exclusively in males, the possibilities are $r_mn\phi/(\mu+n)$, $n=1,\ldots,[1/(a+d)]_i$. Observing that $n/(\mu+n)$ is an increasing function of $n$, the optimal return is $r_m[1/(a+d)]_iS_P$. (This last fact remains true if $S_P$ is replaced by $S$.) Thus, one need only compare these two expressions to decide whether it is more optimal to invest in females or males.

Now suppose $(a+d)+(\alpha+\delta) \leq 1$, equivalently, $a+d \leq 1-\alpha-\delta$, so that investment in both genders is possible. First consider Model 2:
$$R_2(x) = r_m\left[{x \over a+d}\right]_iS + r_f\left[{1-x \over \alpha+\delta}\right]_i.\eqno(5.4)$$
One can analyse this situation conceptually. Suppose
$$1 = p(a+d) + \epsilon_1,\qquad p\hbox{ a nonnegative integer},\qquad 0 \leq \epsilon_1 < (a+d).\eqno(5.5)$$
On each subinterval of the form $\bigl[s(a+d),(s+1)(a+d)\bigr)$, $s = 0,\ldots,p-1$, and on $\bigl[p(a+d),1\bigr]$, ${\cal M}$ is constant and given by ${\cal M}\bigl(s(a+d)\bigr)$ while ${\cal F}$ is nonincreasing. Thus, $R_2$ assumes its largest value on each of these subintervals at the initial point of the subinterval (though not necessarily uniquely so). It follows that the optimal strategy for $R_2$ occurs at one of:
$$x = s(a+d),\qquad s = 0,\ldots,p.\eqno(5.6)$$
Note that there may be other values of $x$ at which an equally optimal strategy occurs. Such values arise because neither $\cal M$ nor $\cal F$ may change for small changes in $x$. For example, the analogous approach to (5.7) based on increments of $\alpha+\delta$ would proceed in such increments down from 1 towards 0. Note that the resulting points $y = 1 - \ell(\alpha+\delta)$ may not have any points in common with (5.7). If not, this fact is explained by the previous remark. As long as one finds some $x$ that gives the optimal strategy, it is a matter of convenience what form one takes; unless there are restrictions on the number of males or females available for translocation. Here I take the form given by (5.7); but the second approach will prove convenient in the next section for case studies with step-function females and male returns of other forms.

If $a+d \geq \alpha+\delta$, write
$$a+d = q(\alpha+\delta) + \epsilon_2,\qquad q\hbox{ a nonnegative integer},\qquad 0 \leq \epsilon_2 <\alpha+\delta.\eqno(5.7)$$
To compute the difference in return $R_2$ between $x=s(a+d)$ and $x=(s+1)(a+d)$, suppose
$$1-s(a+d) = \ell_s(\alpha+\delta) + \epsilon_s,\qquad \ell_s\hbox{ a nonnegative integer},\qquad 0 \leq \epsilon_s < \alpha+\delta.\eqno(5.8)$$
Then
$$R_2\bigl(s(a+d)\bigr) = sr_mS + r_f\ell_s$$
while
$$R_2\bigl((s+1)(a+d)\bigr) = (s+1)r_mS + r_f\left[{1-(s+1)(a+d) \over \alpha+\delta}\right]_i.$$
By assumption $0 \leq 1-(s+1)(a+d)$, whence $1-(s+1)(a+d) + (a+d) = 1 - s(a+d)$ yields $\ell_{s+1}(\alpha+\delta) +\epsilon_{s+1} + q(\alpha+\delta) + \epsilon_2 = \ell_s(\alpha+ \delta) + \epsilon_s$, which entails $\ell_s \geq q$. Hence, $1-(s+1)(a+d) = (\ell_s-q)(\alpha+\delta) + \epsilon_s - \epsilon_2$, and
$$\left[{1-(s+1)(a+d) \over \alpha+\delta}\right]_i = \cases{\ell_s-q,&if $\epsilon_s \geq \epsilon_2$;\cr \ell_s-q-1,&if $\epsilon_s < \epsilon_2$.\cr}$$
Thus,
$$R_2\bigl((s+1)(a+d)\bigr) - R_2\bigl(s(a+d)\bigr) = r_mS - \cases{r_fq,& if $\epsilon_s \geq \epsilon_2$;\cr r_f(q+1),& if $\epsilon_s < \epsilon_2$,\cr}$$
i.e., 
$$R_2\bigl((s+1)(a+d)\bigr) \geq R_2\bigl(s(a+d)\bigr)\ \Leftrightarrow\ {r_mS \over r_f} \geq \cases{\left[{a+d \over \alpha+\delta}\right]_i,& if $\epsilon_s \geq \epsilon_2$;\cr \cr\left[{a+d \over \alpha+\delta}\right]_i+1,& if $\epsilon_s < \epsilon_2$,\cr}.$$
Hence, with $a+d \geq \alpha+\delta$, when
$${r_mS \over r_f} \geq \left[{a+d \over \alpha+\delta}\right]_i + 1,\eqno(5.9a)$$
it does not diminish returns to divert a further $a+d$ resources from females to males, i.e., (5.9a) implies the optimal return occurs at $x = p(a+d)$, i.e., invest in as many males as possible and the residual $1-p(a+d)$ in females if possible. If, on the other hand,
$${r_mS \over r_f} \leq \left[{a+d \over \alpha+\delta}\right]_i,\eqno(5.9b)$$
it never pays to divert $a+d$ from females to males, so the optimal return is obtained by investing all resources in females. Any left over resources will be less than $\alpha+\delta$, whence less than $a+d$, and therefore insufficient for any male. Finally, if
$$\left[{a+d \over \alpha+\delta}\right]_i + 1 > {r_mS \over r_f} > \left[{a+d \over \alpha+\delta}\right]_i,\eqno(5.9c)$$
then at any given $x=s(a+d)$ it may or may not pay to divert a further $a+d$ in resources from females to males. In this case, one must examine each possibility in (5.6).

Now suppose instead that $a+d < \alpha+\delta$. One can analyse this case by suitably interchanging the roles of males and females.  Let ${\cal M}'$ be the function (5.2) that results from the female model parameters $\alpha$, $\delta$, $r_f$ and ${\cal F}'$ the function (5.3) that results from the male model parameters $a$, $d$, $r_m$. Then ${\cal M}'(x) = {\cal F}(x)$ and ${\cal F}'(z) = {\cal M}(z)$. Consider ${\cal M}'(y)(1/S) + {\cal F}'(1-y)$. Since $S$ is a constant, optimization of this expression wrt $y$ is equivalent to optimizing ${\cal M}'(y) + {\cal F}'(1-y)S$ wrt $y$, i.e., equivalent to optimizing ${\cal F}(y) + {\cal M}(1-y)S$ wrt $y$, which is equivalent to optimizing ${\cal M}(x)S + {\cal F}(1-x)$ wrt $x=1-y$. Hence, interchanging the roles of males and females and inverting the sex ratio yields a problem that is equivalent to the original model. Thus, one solves the transformed model as in the previous paragraph; if the solution is to apply resources $y$ to `males' represented by ${\cal M}'$, then the actual optimal solution is to apply resources $y$ to actual females.

For a direct approach to the case when $a+d < \alpha+\delta$, the question is whether it pays, at $x=s(a+d)$, to divert a further $a+d$ in resources from females to another male. The gain from the additional male is $r_mS$. Again write $1-s(a+d) = \ell_s(\alpha+\delta) + \epsilon_s$. Since $a+d < \alpha+\delta$, there is no loss in return from females if $a+d \leq \epsilon_s$ and a loss of $r_f$ if $a+d > \epsilon_s$ (one loses at most one female in this case to gain $a+d$ for a male). In the latter case, the net change is $r_mS - r_f$. Whether the latter case occurs depends on $\epsilon_s$, which depends on the model parameters. In any case, one can assert that if 
$$a+d < \alpha+\delta\qquad\hbox{and}\qquad{r_mS \over r_f} \geq 1\eqno(5.10a)$$
then it does not diminish returns $R_2$ to divert further resources from females to males, whence the optimal return is obtained by investing in as many males as possible (the residue is too small to invest in a female). On the other hand, if
$$a+d < \alpha+\delta\qquad\hbox{and}\qquad{r_mS \over r_f} < 1\eqno(5.10b)$$
then it will not pay to divert $a+d$ to another male if doing so reduces the number of females invested in (by one). In this case, one can only say the optimal return occurs at some point of (5.6); computation will determine which.

Observe that if one solves this scenario using the symmetry argument, then (5.9b) asserts that if $r_f/r_mS < [(\alpha+\delta)/(a+d)]_i$ then the optimal strategy is to invest in all ``females'', i.e., in actual males. Since $a+d < \alpha+\delta$, then $[(\alpha+\delta)/(a+d)]_i \geq 1$. Thus, (5.10a) is consistent with (5.9b). If (5.10b) holds, any of (5.9a--c) may in principle hold.

Note that the conditions (5.9--10) only involve the ratio of $r_mS$ to $r_f$ and the ratio of $a+d$ to $\alpha+\delta$.

It is a simple matter to compute $R_2$ for specified values of $x$, in particular the values (5.6), with the model parameters as variables (for example in MS Excel) so that one can easily numerically solve the model and generate examples confirming (5.9--10). For example, with $\alpha = 0.05$, $\delta = 0.1$, $a = 0.2$, $d = 0.05$, $r_f = 1$, $r_m = 1.5$, $S= 1$, then $1 < (a+d)(\alpha+\delta)$ and $[(a+d)(\alpha+\delta)]_i < r_m/r_f < [(a+d)(\alpha+\delta)]_i + 1$. Consistent with (5.9c), one finds the optimal strategy occurs at $x=0.25$, i.e., $s=1$ in (5.6), so a mixed strategy of investing in 1 male and 5 females yields the optimal return for the given model parameters.

Now consider Model 1 under the assumption $(a+d)+(\alpha+\delta) \leq 1$; equivalently, $a+d \leq 1-\alpha-\delta$. Then, $R_1(x)$ is obtained by maximizing
$$r_m n{\phi+k \over \mu + n} + r_f k,\eqno(5.11a)$$
with respect to $n$ and $k$ subject to
$$0 \leq n \leq \left[x \over a+d \right]_i \hskip 1.25in 0 \leq k \leq \left[1-x \over \alpha+\delta\right]_i,\eqno(5.11b)$$
but where $n=0$ only on $[0,a+d)$ and $k=0$ only on $(1-(\alpha+\delta),1]$. As a function of $n$, the first summand is increasing and so is maximal for $n = [x/(a+d)]_i$. With this choice, (5.11a) is also an increasing function of $k$, whence maximal for $k = [(1-x)/(\alpha+\delta)]_i$. Hence,
$$R_1(x) = {\cal M}(x)S_x + {\cal F}(1-x) = R_3(x),\eqno(5.12)$$
i.e., (4.1) reduces to (4.3); Model 1 reduces to Model 3.

So consider Model 3. Recalling (5.5), on a subinterval $\bigl[s(a+d),(s+1)(a+d)\bigr)$, $s=0,\ldots,p-1$ or $\bigl[p(a+d),1\bigr]$, since 
$$R_3(x) = r_m\left[{x \over a+d}\right]_i\left({\phi+\left[{1-x \over \alpha+\delta}\right]_i \over \mu+\left[{x \over a+d}\right]_i}\right) + r_f\left[{1-x \over \alpha+\delta}\right]_i,$$
then as $[(1-x)/(\alpha+\delta)]_i$ is nonincreasing and $[x/(a+d)]_i$ is constant, then ${\cal M}(x)$ is constant, ${\cal F}(1-x)$ is nonincreasing, and $S_x$ is also nonincreasing (the denominator is constant and the numerator nonincreasing). Hence, $R_3(x)$ takes its maximum value on each of these subintervals at the initial point of the subinterval (though not necessarily uniquely). Hence, as for $R_2$, one can find an optimal strategy from amongst the possibilities (5.6). This observation limits the numerical analysis required. Note, however, that the gain in diverting $a+d$ from females to another male is offset not only by the loss of the return from the females but also by the decrease in $S_x$ through the numerator, so that (5.9) is not valid for Model 1.

Suppose $\alpha = 0.05$, $\delta = 0.05$, $a = 0.2$, $d = 0.025$, $r_m = 4$, $r_f = 1$, $\phi = 10$, $\mu = 5$ (whence $S = 2$). Then $(a+d)/(\alpha+\delta) > 1$, $Sr_m/r_f = 8 > [(a+d)/(\alpha+\delta)]_i + 1 = 3$, so in Model 2 (5.9a) predicts the optimal strategy would be to invest in as many males as possible. Indeed, for Model 2, $x=0.9$, i.e., $s=4$ in (5.6), is the optimal strategy, investing in 4 males and 1 female for a return of 33. For Model 1, however, one computes that the optimal return is 22.5 and can be achieved with $s=3$ in (5.6) to give $x = 0.675$, i.e., by investing in 3 males and 3 females. In Model 1, the strategy $x = 0.9$ yields a return of 20.5.
\vskip 24pt
\noindent {\section 6. STEP-FUNCTION FEMALES AND NON-STEP-FUNCTION MALES}
\vskip 12pt
In this section I assume the individual female return function is as in (5.1) while the individual male return function is assumed to be a nondecreasing function (other than a step function) with a packaging cost of $d$. I restrict attention to the case for which $d + (\alpha+\delta) < 1$, equivalently $d < 1 - (\alpha+\delta)$, so that investment in both genders is possible. For Model 2, one therefore has (4.2) with ${\cal F}(z)$ as in (5.3). Write
$$1 = n(\alpha+\delta) + \epsilon,\qquad n \hbox{ a nonnegative integer},\qquad 0 \leq \epsilon < \alpha + \delta\eqno(6.1)$$
and choose the nonnegative integer $k$ so that
$$w := 1-k(\alpha+\delta) > d\qquad\hbox{but}\qquad 1 - (k+1)(\alpha+\delta) \leq d,\eqno(6.2)$$
i.e., subtracting increments of $\alpha+\delta$ from 1, $w$ is the last quantity so obtained that is greater than $d$. Observe that, on each of the subintervals
$$\bigl(1-(\alpha+\delta),1\bigr],\ \bigl(1-2(\alpha+\delta),1-(\alpha+\delta)\bigr],...,\bigl(d,1-k(\alpha+\delta)\bigr],\eqno(6.3)$$
the function ${\cal F}(1-x)$ is constant, given by its value at the right-hand endpoint.
\vskip 12pt
\noindent {6.4 Lemma}\hfil\break
Whenever $m(x)$ is a nondecreasing function of $x$, so is ${\cal M}(x)$. 

Proof: For resources $y > x$, one can of course obtain the return ${\cal M}(x)$ with investment of resources $x$, leaving $y-x$ further to invest, so one can do no worse than ${\cal M}(x)$, i.e., ${\cal M}(y) \geq {\cal M}(x)$. For $m$ as in (3.d) or (3.e), for example, this result is explicitly of the form: ${\cal M}(y) = n_ym(y/n_y)$, which is at least as large as $n_xm(y/n_x)$ since $n_y$ is by definition the optimal equitable division of resources $y$; but since $m$ is nondecreasing, $n_xm(y/n_x) \geq n_xm(x/n_x) = {\cal M}(x)$.
\vskip 12pt
It follows from (6.4) and the constancy of ${\cal F}(1-x)$ on each of the subintervals of (6.3) that $R_2(x)$ is nondecreasing on each of these subintervals, whence its maximal value on each subinterval of (6.3) occurs at the right-hand endpoint. Moreover, since $R_2(x) = F(1-x)$ on $[0,d]$, then $R_2(x)$ is nonincreasing on [0,d]. Thus, the optimal return in Model 2, whenever females have step-function return, occurs at one of the points:
$$1,\ 1-(\alpha+\delta),\ 1-2(\alpha+\delta),\ldots,1-k(\alpha+\delta)=w,\ 0.\eqno(6.5)$$
For Model 2, it therefore suffices to compute the return $R_2$ for this finite list of possibilities and select the maximum value to determine the optimal return.

Now consider Model 1. From (4.1) and (5.1), it is clear that the optimal strategy requires investment of $1-x$ in $k_{1-x} := \left[(1-x)/(\alpha+\delta)\right]_i$ females so that $R_1$ takes the form
$$R_1(x) = P(x,n_x)(\phi+k_{1-x}) + r_fk_{1-x},\eqno(6.6)$$
where $P(x,n) := nm(x/n)/(\mu+n)$ and $n_x$ is the nonnegative integer maximizing $P(x,n)$ subject to (A.2) ($n_x=1$ for 3.b--c). Hence, if $y > x$, because $P(y,n_y) \geq P(y,n)$ for any $n$ satisfying (A.2),
$$P(y,n_y) := {n_ym(y/n_y) \over \mu + n_y} \geq {n_xm(y/n_x) \over \mu + n_x} \geq {n_xm(x/n_x) \over \mu + n_x} = P(x,n_x).$$
Thus, $P(x,n_x)$ is a nondecreasing function of $x$ and the remaining terms of (6.6) are constant on the subintervals (6.3), so $R_1(x)$ is nondecreasing on these subintervals and thus takes its maximal value on each subinterval at the right-hand endpoint. As $R_1(x) = {\cal F}(1-x)$ on $[0,d]$, then as with Model 2, it suffices to compute the return $R_1$ for the finite list of possibilities (6.5) and select the maximum value to determine the optimal return.

I present details of specific cases in the following subsections.
\vskip 12pt
\noindent {\subsec 6.a Linear Males}
\vskip 12pt
The individual return function for males is given by (3.b.1). As usual, I suppose $d+(\alpha+\delta) < 1$, equivalently, $d < 1-\alpha-\delta$, in order that investment in both genders is possible. 

For Model 2, one has
$$R_2(x) = \cases{r_f\left[{(1-x) \over \alpha+\delta}\right]_i,& if $x \in [0,d]$;\cr
\cr
a(x-d)S + r_f\left[{1-x \over \alpha+\delta}\right]_i,& if $x \in [d,1]$.\cr}\eqno(6.{\rm a}.1)$$
I will derive a complete characterization of the optimal strategy in terms of the model parameters.

Assume that 
$$a(\alpha+\delta)S > r_f,\eqno(6.{\rm a}.2)$$
i.e., the marginal return on investing $\alpha+\delta$ in a male (beyond the initial investment of $d$) exceeds the marginal return from investment of the same resources in a single female. Then,
$$R_2(1) > R_2(1-\alpha-\delta) > \cdots > R_2\bigl(1-(k-1)(\alpha+\delta)\bigr) > R_2(w)\eqno(6.{\rm a}.3)$$
since each term is the previous term plus $-a(\alpha+\delta)S + r_f$. Thus, granted (6.a.2), (6.5) implies that $R_2$ has its maximum value on $(d,1]$ at $x=1$. It then follows from (6.5) that the larger of $R_2(1)$ and $R_2(0)$ is the maximum value of $R_2$.

If, on the other hand,
$$a(\alpha+\delta)S_P < r_f,\eqno(6.{\rm a}.4)$$
then
$$R_2(1) < R_2(1-\alpha-\delta) < \cdots < R_2\bigl(1-(k-1)(\alpha+\delta)\bigr) < R_2(w),\eqno(6.{\rm a}.5)$$
whence from (6.5) the maximum value of $R_2$ on $(d,1]$ occurs at $x=w$. Hence, by (6.5), the maximum value of $R_2$ occurs at $x=w$ or $x=0$ (the return on investing $\alpha+\delta$ in another female is greater than the marginal return on the same investment in a male, so the optimal strategy is to invest in as many females as possible, either exclusively females or the maximum number with enough left over to invest in a male with return). Clearly, if $R_2(0) \leq R_2(1)$, then as $R_2(1) < R_2(w)$, then the optimal strategy occurs at $x=w$.

To determine the optimal strategy, otherwise, i.e., when (6.a.4) holds and $R_2(0) > R_2(1)$, recall (6.1), and suppose first that $d < \epsilon$, whence the $k$ of (6.2) equals $n$ and $w = \epsilon$. In fact, irrespective of the relative magnitudes of $R_2(0)$ and $R_2(1)$, when $d < \epsilon$,
$$R_2(0) = r_fn < r_fn + a(\epsilon -d)S = R_2(\epsilon) = R_2(w),$$
i.e., $x=0$ is not optimal.

Finally, suppose instead that $d \geq \epsilon$, whence the $k$ in (6.2) is less than $n$. Now, using (6.1),
\vskip 12pt
$$\eqalign{R_2(w) &= r_fk + a\bigl(1-k(\alpha+\delta)-d\bigr)S\cr
&= r_fk + a(1-d)S - ak(\alpha+\delta)S\cr
&\leq r_fk + a(1-\epsilon)S - ak(\alpha+\delta)S\cr
&= r_fk + an(\alpha+\delta)S - ak(\alpha+\delta)S\qquad\hbox{by (6.1)}\cr
&= r_fk + a(n-k)(\alpha+\delta)S\cr
&< r_fk + (n-k)r_f\qquad\hbox{by (6.a.4)}\cr
&=r_fn = R_2(0).\cr}$$
\vskip 12pt
In summary, the optimal return must occur at $x=1$, $x=0$, or the interior solution $w$ given by (6.2), and these possibilities occur as follows: 
$$\displaylines{R_2(1) > R_2(0)\hbox{ \& }aS(\alpha+\delta) \geq r_f\qquad\hbox{then}\qquad x=1\cr
\hbox{(if $R_2(1) = R_2(0)$, then $x=0$ and $x=1$ give the same return)};\cr
R_2(0) > R_2(1)\hbox{ \& }aS(\alpha+\delta) \geq r_f\qquad\hbox{then}\qquad x=0;\cr
\hfill R_2(1) \geq R_2(0)\hbox{ \& }aS(\alpha+\delta) < r_f\qquad\hbox{then}\qquad x=w;\hfill\llap(6.{\rm a}.6)\cr
\noalign{\vskip 6pt}
R_2(0) > R_2(1)\hbox{ \& }aS(\alpha+\delta) < r_f\hbox{ \& }\cases{d < \epsilon,&then $x=w$;\cr
d \geq \epsilon,& then $x=0$.\cr}\cr}$$
Each of these scenarios is realizable with appropriate choices of model parameters.

Now consider Model 1. Investing in more than a single male decreases the return from males both through packaging costs (as in 3.b) and through the weighting factor $S_P$. Hence, Model 1 reduces to Model 3 and (6.6) becomes
$$R_1(x) = \cases{r_f\left[{1-x \over \alpha+\delta}\right]_i,&if $x \in [0,d]$;\cr
a(x-d)\left({\phi+\left[{1-x \over \alpha+\delta}\right]_i \over \mu+1}\right) + r_f\left[{1-x \over \alpha+\delta}\right]_i,& if $x \in [d,1]$;\cr}\eqno(6.{\rm a}.7)$$
which is in fact continuous at $x=d$.

The optimal return must occur at one of (6.5) but the varying value of $S_P$ now complicates matters. One has:
$$\displaylines{R_1(1) = a(1-d){\phi \over \mu+1};\cr
R_1\bigl(1-(\alpha+\delta)\bigr) = r_f + a\bigl(1-(\alpha+\delta)-d\bigr){\phi+1 \over \mu+1} = \left[a(1-d){\phi+1 \over \mu+1}\right] + \left[r_f - a(\alpha+\delta){\phi+1 \over \mu+1}\right];\cr
R_1\bigl(1-2(\alpha+\delta)\bigr) = 2r_f + a\bigl(1-2(\alpha+\delta)-d\bigr){\phi+2 \over \mu+1} = \left[r_f+a\bigl(1-(\alpha+\delta)-d\bigr){\phi+2 \over \mu+1}\right] + \left[r_f - a(\alpha+\delta){\phi+2 \over \mu+1}\right];\cr
\hfill\vdots\hfill\llap(6.{\rm a}.7)\cr
R_1(w)=R_1\bigl(1-k(\alpha+\delta)\bigr) = kr_f + a\bigl(1-k(\alpha+\delta)-d\bigr){\phi+k \over \mu+1}\cr 
= \left[(k-1)r_f+a\bigl(1-(k-1)(\alpha+\delta)-d\bigr){\phi+k \over \mu+1}\right] + \left[r_f - a(\alpha+\delta){\phi+k \over \mu+1}\right];\cr
R_1(0) = r_fn.\cr}$$
For $r=1,\ldots,k$, consider the final expression for $R_1\bigl(1-r(\alpha+\delta)\bigr)$. The first bracketed quantity is larger than $R_1\bigl(1-(r-1)(\alpha+\delta)\bigr)$. The second summand can be either positive or negative. Thus, depending on these values, it seems likely that the optimal return could occur at any one of (6.5), yielding a more complicated situation than for $R_2$. 

Consider two possible simplifications. Suppose
$$r_f > a(\alpha+\delta){\phi+k \over \mu+1} > a(\alpha+\delta){\phi+(k-1) \over \mu+1} > \cdots > a(\alpha+\delta){\phi+1 \over \mu+1},\eqno(6.{\rm a}.8)$$
noting that the only assumption here is the first inequality. Then from (6.a.7) one sees that
$$R_1(w) > R_1\bigl(1-(k-1)(\alpha+\delta)\bigr) > \cdots > R_1\bigl(1-(\alpha+\delta)\bigr) > R_1(1),\eqno(6.{\rm a}.9)$$
whence the maximum of $R_1$ on $(d,1]$ occurs at $x=w$. It follows that the maximum of $R_1$ occurs at $x=w$ or $x=0$ if (6.a.8) holds. This result is an analogue of the results obtained above for $R_2$ under the assumption that $r_f > a(\alpha+\delta)S$.

If on the other hand
$$r_f < a(\alpha+\delta){\phi+1 \over \mu+1} < a(\alpha+\delta){\phi+2 \over \mu+1} < \cdots < a(\alpha+\delta){\phi+k \over \mu+1},\eqno(6.{\rm a}.9)$$
where again only the first inequality is an assumption, the second summand in each line of (6.a.7) is now negative, but whether sufficiently negative to make $R_1\bigl(1-r(\alpha+\delta)\bigr)$ smaller than $R_1\bigl(1-(r-1)(\alpha+\delta)\bigr)$ is undetermined, which reinforces the possibility that any of (6.5) might give the optimal return, depending on the values the model parameters take.

In any event, one need only examine the possibilities in (6.5) to find the optimal strategy.

Consider $\alpha = 0.05$, $\delta = 0.05$, $r_f = 1.5$, $a = 10$, $d=0.05$, $S=3$ whence (6.a.6) is satisfied. With $S_P = 30/10$ there is no difference between Model 2 and 1 as regards optimal strategy: choose $x=1$, i.e., invest all resources in a single male. But, in Model 1, putting $S_P = 6/2$, which involves no change for Model 2, the optimal strategy for Model 1 is now: choose $x=0.6 = 1 - 4(\alpha+\delta)$, which gives a return of 24.3, by investing only 60\% of resources in the male and the remaining 40\% equitably amongst 4 females, versus a return of 19 for $x=1$. Moreover, $x=0.6$ is not a possible optimal strategy for Model 2 at all ($w = 0.1$). This example underscores the difference between Models 1 and 2.
\vskip 12pt
\noindent {\subsec 6.b DMR Males}
\vskip 12pt
With $f$ as in (5.1) and $m$ as in (3.d.1), one can write for Model 2
$$R_2(x) = n_xm\left({x \over n_x}\right)S + r_f\left[{1-x \over \alpha+\delta}\right]_i,\eqno(6.{\rm b}.1)$$
where, as usual, $n_x$ is the optimal number of males in which to invest resources $x$, with $n_x = 0$ understood for $x \in [0,d]$ so the first term reduces to zero on $[0,d]$. Assume that $d + \alpha+\delta < 1$ so that investment in both genders is possible. 

Suppose $m$ is known explicitly. If one can solve (3.d.7) explicity for $t$, $n_x$ can be determined from (3.d.8); otherwise, one can approximate $t$ with sufficient accuracy for numerical computations. For certain return functions, one can compute $\cal M$ from the observations in Appendix B. Alternatively, one can compute the quantity $nm(x/n)$ for the allowed values (A.2) of $n$ and determine $n_x$ by inspection. Thus, for an explicit $m$, one can compute $\cal M$, and then $R_2$, at each of the values in (6.5) and numerically determine the optimal return in Model 2. Can one say more?

If one employs the approximation
$$n_x = {x \over t},\eqno(6.{\rm b}.2)$$
with $t$ the solution of (3.d.7), then in effect the male return function becomes piecewise-linear and one obtains an analytic expression for the resulting approximation to $R_2$:
$$R^a_2(x) := \cases{r_f\left[{z \over \alpha+\delta}\right]_i,& if $x \in [0,d]$;\cr
(xm(t)/t)S + r_f\left[{z \over \alpha+\delta}\right]_i,& if $x \in (d,1]$.\cr}\eqno(6.{\rm b}.3)$$
This expression is of the form ${\cal M}^a(x)S + {\cal F}(1-x)$ with
$${\cal M}^a(x) := \cases{0,& if $x \in [0,d]$;\cr xm(t)/t=xm'(t),& if $x \in (d,1]$;\cr}\eqno(6.{\rm b}.4)$$
noting (3.d.7). Thus, (6.b.3) differs only from the form of (6.a.1) in that ${\cal M}^a(x) \propto x$ rather than $(x-d)$, which makes ${\cal M}^a$ discontinuous at $x=d$ (note that the packaging costs $d$ have the effect of making the approximation for male returns piecewise linear). Working through the argument of \S 6.a as applied to (6.b.3), one obtains a modified version of (6.a.6):
$$\displaylines{R_2^a(1) > R_2^a(0)\hbox{ \& } m'(t)(\alpha+\delta)S \geq r_f\qquad\hbox{then}\qquad x=1\cr
\hbox{(if $R_2^a(1) = R_2^a(0)$, then $x=0$ and $x=1$ give the same return)};\cr
\hfill R_2^a(1) \geq R_2^a(0)\hbox{ \& }m'(t)(\alpha+\delta)S < r_f\qquad\hbox{then}\qquad x=w;\hfill\llap(6.{\rm b}.5)\cr
\noalign{\vskip 6pt}
R_2^a(0) > R_2^a(1)\hbox{ \& }m'(t)(\alpha+\delta)S < r_f\hbox{ \& }\cases{d < \epsilon,&then $x=w$;\cr
d \geq \epsilon,& then $x=0$ or $w$.\cr}\cr}$$
Because $R_2^a(1) = m'(t)S$ while $R_2^a(0) = r_f\left[1/(\alpha+\delta)\right]_i \leq r_f/(\alpha+\delta)$, $r_f \leq m'(t)(\alpha+\delta)S$ entails $R_2^a(0) \leq R_2^a(1)$ and the second possibility in (6.a.6) will not in fact occur here. But, while 
$$m'(t)S \geq {r_f \over \alpha + \delta} \geq {r_f(1-\epsilon) \over \alpha+\delta} = R_2(0)$$
$m'(t)S \geq R_2(1)$ so in the actual model $R_2(1) < R_2(0)$ may not be ruled out when $m'(t)(\alpha+\delta)S \geq r_f$. Moreover, in place of (6.a.4), one only finds $R^a_2(w) < R^a_2(0) + m'(t)S\epsilon$, so the final case in (6.a.6), is here left unresolved, the condition $ d \geq \epsilon$ being uninformative; one must simply compute the two possibilities.

Note that this approximation, if useful at all, is useful for obtaining the value of $x$ for which $R_1$ or $R_2$ is maximized. The optimal strategy must still be interpreted as equitable partition of resources $x$ in $n_x$ males, with $n_x$ given by (3.d.8) (and equitable partition of $1-x$ in females of course). If the optimal return occurs in this approximation at $x=1$, then of course $n_x = [1/t]_i$.

How useful is the approximation (6.b.2) and its consequence (6.b.5)? The error in (6.b.2) is less than one, i.e., $\vert n_x -x/t \vert < 1$. Since $v_x := x/t$ is the actual value at which $\chi_x(v) = vm(x/v)$ achieves its maximum (\S 3.d), then the approximation (6.b.2) overestimates the actual return on the investment in males. The error will therefore be magnified by $S >1$ and diminished by $S < 1$. Thus, when (6.b.5) favours investment in females over males, that deduction should be robust. When (6.b.5) favours investment in males over females, that deduction may be questionable if the overestimation of male returns is significant. 

In various trials with $m = g$ as in (3.d.10) (especially with $s=1$ for which $\chi_x(v)$ is quadratic) the linear approximation (6.b.2--5) successfully predicted the optimal strategy. But, with $d = 0.07$, $K = 5$ and $s = 1$ in (3.d.10) and with $\delta = 0.04$, $\alpha = 0.03$, and $r_f = 1.4$ in (5.1), then  $m'(t)(\alpha+\delta)S < r_f$ and $R^a_2(1) < R^a_2(0)$ so (6.b.5) implies the optimal return occurs at $x=w = 0.02$ or $x=0$. In fact, $R_2^a$ has its maximum value of 19.81 at $x=w$ (though $R_2^a(0) = 19.6$, only slightly less) while $R_2(w) = 19.31$ which is less than $R_2(0) = 19.6$, the actual maximal return being $R_2(0.16) = 19.61$. So, in this case, the approximation is slightly misleading. Note that $0.16 = 1-12(\alpha+\delta)$, so this result is consistent with (6.5). It may well be that the difference in returns between the strategy suggested by (6.b.5) and the true optimal strategy, as in the example just presented, is of no practical significance. For an explicit choice of function for $m$, one can compute the error involved in utilizing $R_2^a$, but since the optimal return for $R_2$ can be obtained readily from (6.5) the approximation (6.b.2--5) is mostly of theoretical interest in suggesting that Model 2 with DMR males behaves similarly to Model 2 with linear males.

Now consider Model 1. Since for a fixed number of males invested in, the optimal partition of resources is equitable distribution, for any $x \in [0,1]$, the expression to be optimized to obtain $R_1$ is
$$nm(x/n)\left({\phi+k \over \mu + n}\right) + r_fk\qquad 0 \leq n < {x \over d},\qquad 0 \leq k \leq \left[{1-x \over \alpha+\delta}\right]_i,\eqno(6.{\rm b}.6)$$
where $n=0$ on, and only on, $[0,d]$ and $k=0$ on, and only on, $(1-\alpha-\delta,1]$. It is clear that $k$ should take its maximal value. On $(d,1]$, therefore, (6.b.6) becomes
$$nm(x/n)\left({\phi+\left[{1-x \over \alpha+\delta}\right]_i \over \mu + n}\right) + r_f\left[{1-x \over \alpha+\delta}\right]_i\qquad 0 \leq n < {x \over d},\eqno(6.{\rm b}.7)$$
in agreement with (6.6), and where for $x \in [0,d]$ it is understood that the first summand vanishes. 

The final optimization required to obtain $R_1$ is with respect to $n$. Analogous to the procedure employed to obtain $n_x$ in \S 3.4, for $x \in (d,1]$, consider the function
$$\psi_x(v) := vm(x/v)\left({k_x \over \mu +v}\right),\eqno(6.{\rm b}.8)$$
defined on the interval $(0,x/d)$, where $k_x = \phi + [(1-x)/(\alpha+\delta)]_i$. We aim to find the maximal value of $\psi_v$ as a function of $v$ for each $x \in (d,1]$ and deduce therefrom the optimal $n_x$. By direct calculation, and after some simplification:
$$\psi'_x(v) = {k_x \over \mu+v}\left[m\left({x \over v}\right){\mu \over \mu+v}-m'\left({x \over v}\right)\left({x \over v}\right)\right];\eqno(6.{\rm b}.9)$$
$$\psi''_x(v) = {-2k_x \over (\mu+v)^2}\psi'_x(v) + {k_x \over \mu+v}m''\left({x \over v}\right){x^2 \over v^3};\eqno(6.{\rm b}.10)$$
which is not as simple as was obtained for $\phi_x$ in \S 3.4. First note that $\psi'_x$ is zero when
$$m\left({x \over v}\right){\mu \over \mu+v} = m'\left({x \over v}\right)\left({x \over v}\right).\eqno(6.{\rm b}.11)$$
If $v_x$ is a solution of this equation, with $\tau_x := x/v_x$, then (6.b.11) becomes
$${m(\tau_x) \over \tau_x}{\mu \over \mu + x/\tau_x} = m'(\tau_x).\eqno(6.b.12)$$
Interpreting this equation geometrically, since $\mu/(\mu + x/\tau_x) < 1$, $\tau_x$ is a point where the line through the origin and the point $\left(\tau_x,m(\tau_x)\right)$ is steeper than the tangent line to $m$ through the same point. From the shape of $m$, it is then clear that $\tau_x > t$. From (6.b.10), $\psi''_x(v_x) < 0$ since $m''(\tau_x) <0$, i.e., each stationary point of $\psi_x$ is indeed a relative maximum. Observe, from (6.b.10), that $\psi''_x(v)$ is negative wherever $\psi'_x(v) > 0$, but that when $\psi'_x(v) < 0$, $\psi''_x(v)$ may become positive. In particular, assuming $\psi_x$ is $C^2$, as $v$ increases away from $v_x$ one expects $\psi''_x$ may eventually change sign. But since there are no relative minima, a change in concavity cannot result in $\psi_x$ exceeding $\psi_x(v_x)$ whence there can only be the one relative maximum. At $v = x/d$ observe that (6.b.8) vanishes, whence, using (6.b.8--10), $\psi_x$ can be extended to $(0,x/d]$ in a $C^2$ fashion. Since one is only interested in the restriction of $\psi_x$ to $[1,x/d]$, on which $\psi_x$ is bounded and positive, and $\psi_x$ has a unique relative maximum on $(0,x/d)$ and no other critical points, its maximum on $[1,x/d]$ must be at $v_x$, whence $n_x$ must be one of the integers either side of (the unique solution to (6.b.11)) $v_x$. Thus, one does obtain a result analogous to (3.d.8).

But, we do not have an explicit characterization of $\tau_x$, only the implicit characterization of (6.b.12) and we do not know the dependence of $\tau_x$ upon $x$. To explore these issues, rewrite (6.b.12) in the form
$${m(\tau_x) - m'(\tau_x)\tau_x \over m'(\tau_x)} = {x \over \mu}.\eqno(6.{\rm b}.13)$$
When $\mu \gg 1$, the right-hand side is approximately zero, whence the numerator of the left-hand side is approximately zero, i.e., $\tau_x \approx t$. Note that $x/\mu$ is a simple linear increasing function of $x$. The expression $(m(z) - m'(z)z)/m'(z)$, as a function of $z$, has derivative $-mm''/(m')^2$, which is indeed positive. Thus, as $x$ increases, the left-hand side of (6.b.13) must increase, and since its a strictly increasing function of $\tau_x$, it must be the case that $\tau_x$ increases with $x$. Since $\tau_x > t$, then as $x$ increases, $m'(\tau_x)$ decreases more rapidly than $m(\tau_x)/\tau_x$. It follows from (6.b.12), that $v_x = x/\tau_x$ increases with $x$.

In summary, one finds
$$R_1(x) = n_xm(x/n_x)\left({\phi+\left[{1-x \over \alpha+\delta}\right]_i \over \mu + n_x}\right) + r_f\left[{1-x \over \alpha+\delta}\right]_i,\eqno(6.{\rm b}.14)$$
where, for $x \in (d,1]$, $n_x$ is the integer either side of $v_x = x/\tau_x$ that maximizes the first summand in (6.b.13), while $n_x = 0$ for $x \in [0,d]$. We have seen that $\tau_x > t$  and that both $\tau_x$ and $v_x$ are increasing functions of $x$. Since $\tau_x > t$, then $v_x := x/\tau_x < x/t$, i.e., not surprisingly, in Model 1, $v_x$, whence possibly $n_x$, is smaller than in Model 2, since the value of males is discounted by their own numbers. Note that Model 1 has not reduced to Model 3. In any event, one computes $R_1$ for each of the possibilities in (6.5) to determine the optimal return; for each such $x$, if necessary, one can compute $nm(x/n)/(\mu+n)$ for the allowed values (A.2) of $n$ and thereby determine $n_x$, thence $R_1(x)$.

Turning to the explicit examples of (3.d.10), for $h$ one finds that (6.b.13) becomes $[(1-s)\tau_x -d]/s = x/\mu$, i.e.,
$$\tau_x = {sx + d\mu \over \mu(1-s)} = t + {sx \over \mu(1-s)}.\eqno(6.{\rm b}.15)$$
As expected, $\tau_x$ is an increasing function of $x$ and for $\mu \gg 1$, $\tau_x \approx t$. Furthermore,
$$v_x = {x \over \tau_x} = {x\mu(1-s) \over sx + d\mu},\eqno(6.{\rm b}.16)$$
which is an increasing function of $x$ with horizontal asymptote at $y=\mu(1-s)/s$. But $x=1$ is the upper limit for our interests, hence $v_x \leq \mu(1-s)/(s+d\mu)$. Thus, for $h$, one has a fairly simple situation.

Now consider $g$. For (6.b13), one obtains
$${\tau_x^{s+1} - (s+1)d^s\tau_x \over sd^s} = {x \over \mu}.\eqno(6.{\rm b}.17)$$
which can be rewritten as the polynomial equation
$$p(z) := z^{s+1} - (s+1)d^sz - (x/\mu)sd^s = 0.\eqno(6.{\rm b}.18)$$
Observe that
$$p(t) = p(\root s \of{s+1}d) = (s+1){\root s \of{s+1}}d^{s+1} - (s+1){\root s \of{s+1}}d^{s+1} - (x/\mu)sd^s = -{x \over \mu}sd^s < 0.$$
By Descartes's rule of signs, (6.b.18) has at most one real positive root. Since $p$ is positive for large $z$ and negative at $z=t$, then $p$ has exactly one positive real root, somewhere on $(t,\infty)$. Also note, 
$$p(1) = 1 - (s+1)d^s - {x \over \mu}sd^s < 1.$$
So, if $p(1) > 0$, then $t < \tau_x \leq 1$. Otherwise, $\tau_x \geq 1$, in which case, as in (3.d.9), $n_x = 1$.

Now $\tau$ (I drop the subscript for convenience) satisfies (6.b.18). Differentiating this equation implicitly with respect to $x$, one finds
$$\tau' = {s \over s+1}{d^s \over \mu}{1 \over \tau^s - d^s},\eqno(6.{\rm b}.19)$$
and as $\tau > d$ then $\tau^s > d^s$ so $\tau'$ is positive, confirming $\tau$ is an increasing function of $x$. As $v_x = x/\tau$, differentiating with respect to $x$ yields,
$$v'_x = {\tau - x\tau' \over \tau^2}.\eqno(6.{\rm b}.20)$$
Substituting (6.b.19) into (6.b.20), the numerator is
$$\tau-x\tau' = {(s+1)\tau^{s+1} - (s+1)d^s\tau - sxd^s/\mu \over (s+1)(\tau^s - d^s)} = {s\tau^{s+1} \over (s+1)(\tau^s - d^s)}$$
since $\tau$ satisfies (6.b.18). As $\tau^s-d^s > 0$, it follows that
$$v'_x = {s\tau^{s-1} \over (s+1)(\tau^s - d^s)} > 0\eqno(6.{\rm b}.21)$$
confirming that $v_x$ is an increasing function of $x$. In general, it does not appear possible to give an explicit formula for the positive real root of (6.b.18), i.e., a formula for $\tau$ as a function of $x$.

Consider the particular case $s=1$, however. Then (6.b.18) is the quadratic
$$z^2 - 2dz - xd/\mu = 0$$
with solutions $d \pm \sqrt{d^2 + xd/\mu}$. The choice of minus sign clearly gives a negative number so
$$\tau_x = d + \sqrt{d^2 +dx/\mu} = t + \sqrt{d^2 +dx/\mu} - d,\eqno(6.{\rm b}.22)$$
since $t=2d$ when $s=1$. Again, $\tau_x \approx t$ when $\mu \gg 1$. Furthermore,
$$v_x = {x \over d + \sqrt{d^2 +xd/\mu}},\eqno(6.{\rm b}.23)$$
and one computes
$$v'_x = {d\sqrt{d^2 +dx/\mu} + d^2 +{xd \over 2\mu} \over (d + \sqrt{d^2 +dx/\mu})^2\sqrt{d^2 +dx/\mu}} > 0,$$
confirming that $v_x$ is an increasing function of $x$ when $s=1$. 

As an example, for $m=g$ with $d = 0.07$, $K = 5.25$, $s=1$, and with $\alpha = \delta = 0.05$, $r_f = 1.5$, and $S = 10/10$, Model 2 has an optimal return of 18.7 for the strategy $x=1$ with equitable partition of all resources amongst 7 males, while Model 1 has an optimal return of 18.4 for the strategy $x = 0.3 = 1 - 7(\alpha+\delta)$ by investing 30\% of resources equitably between 2 males and 70\% equitably amongst 7 females. The return in Model 1 from the strategy $x=1$ is only 11.4.
\vskip 12pt
\noindent {\subsec 6.c Sigmoid Males}
\vskip 12pt
The theoretical discussion for this case is essentially identical to that for DMR males. Note that the consideration of $\psi_x$ is really only relevant when $x > p$ (since $n_x =1$ for $x \in (d,p]$) where $m'' < 0$ as for DMR return functions. Thus, Models 2 \& 1 may be solved as indicated in \S 6.b.

The linear approximation to Model 2, however, is another matter. Recall from \S 3.e that for sigmoid males ${\cal M}(x) = m(x)$, on $[0,p]$. The linear approximation of ${\cal M}$ by ${\cal M}^a(x) = (x/t)m(t)$ cannot be expected to provide a good approximation to $m$ on $[0,p]$. The larger is $p$, the more pronounced will be the disagreement between ${\cal M}$ and ${\cal M}^a$. Of course, the smaller is $p$ the more a sigmoid approximates to a DMR function.

For case studies of sigmoid return functions, I employed a sigmoid return function based on the logistic. Consider, for $x \geq d$,
$$l(x) := {ZL \over (L-Z)e^{-r(x-d)}+Z} - Z,\eqno(6.{\rm c}.1)$$
with $L > Z > 0$, then $l(x)\ \rightarrow\ L-Z$ as $x \rightarrow \infty$ and $l(d) = 0$. The point $p$ of inflexion of $l$ is found to satisfy
$$r = {\ln\left({L-Z \over Z}\right) \over p-d}.\eqno(6.{\rm c}.2)$$
Through this equation, $L$, $Z$, $d$ and $p$ can be taken to be the model parameters (specifying $p$ provides more direct control of whether the sigmoid is more like an IMR or DMR return function, or balanced between these two extremes). Substituting (6.c.2) into (6.c.1) yields
$$l(x) = {ZL \over (L-Z)\left({Z \over L-Z}\right)^{x-d \over p-d}+Z} - Z.\eqno(6.{\rm c}.3)$$
One may compute that
$$l'(p) = {rL \over 4} = {L\ln\left({L-Z \over Z}\right) \over 4(p-d)},\eqno(6.{\rm c}.4)$$
for the maximum rate of change of $l$, and
$$l'(d) = {r(L-Z)Z \over L} = {\ln\left({L-Z \over Z}\right)\left({L-Z \over L}\right)Z \over p-d}.\eqno(6.{\rm c}.5)$$
For $l$ to display its asymptotic behaviour on the interval $(d,1]$, one requires $r$ large, which can be achieved by choosing $Z$ small relative to $p-d$.

Substituting (6.c.1) into (3.4.7) yields
$$rLte^{r(t-d)} = Ze^{2r(t-d)} + (L-2Z)e^{r(t-d)} - (L-Z),$$
which is not analytically solvable for $t$, though one can numerically approximate $t$ for given parameters by plotting $m'(x)x-m(x)$ as a function of $x$ to locate the zero. Thus, the optimal $n$ for any $x$ will have to be determined numerically, either by computing $nm(x/n)$ for the allowed values (A.2) of $n$ or by numerically approximating $t$ and using (3.d.8).

For this return function, the linear approximation to ${\cal M}$ was indeed poor on $[0,p]$. Examination of the function $\phi_x$ for (6.c.1) indicates it is much more peaked than for $g$ and $h$ (and does become concave up towards the right-hand end of its domain) so the approximation of taking $n_x$ to be $v_x = x/t$ is less accurate.

The difference between sigmoid and DMR males is most pronounced, of course, when $p$ is a significant fraction of the available resources. With $L=10$ , $Z= 0.01$, $d=0.05$, $p=0.45$, $\alpha=\delta=0.05$, $r_f = 1.5$ then $w=0.1$. With $S=1$, Model 2 has an optimal return at $x=0.6 = 1-4(\alpha+\delta)$ (this value indicates the poor performance of the linear approximation) achieved by investing 60\% of resources in a single male and the remaining 40\% equitably amongst 4 females. Model 1 achieves its optimal return with the same strategy, independently of the values of $\phi$ and $\mu$ (subject to $\phi/\mu = 1$). If $S = 4$, the optimal strategy becomes $x=1$ for Model 2 and for Model 1 with large populations. When $\mu = 6$ and $\phi=24$, the optimal strategy for Model 1 switches to $x=0.6$ again (investing in one male as before), and remains so for smaller values of $\mu$. This relative stability of the optimal strategy in Model 1 relative to the value of $\mu$ in a fixed ratio $\phi/\mu$ reflects the fact that one must invest more than $p$ in each male to obtain a useful return and here $p$ is a substantial proportion of the available resources, thus severely limiting the number of males one can invest in.
\vskip 24pt
\noindent {\section 7. STEP-FUNCTION MALES AND NON-STEP-FUNCTION FEMALES}
\vskip 12pt
The male return function $m$ is given by (3.a.1), whence $\cal M$ is given by (3.a.2). The packaging cost for females is $\delta$; assume that $a+d+\delta < 1$, equivalently $a+d < 1-\delta$, so that investment in both genders is possible.

Results for Model 2 can be obtained from the results for Model 2 in \S 5.2 as follows. One has
$$R_2(x)= {\cal M}(x)S + {\cal F}(1-x) = n_xr_mS + {\cal F}(1-x),$$
where $n_x = [x/(a+d)]_i$. Substitute $z = 1-x$ and multiply by $1/S$ to obtain
$$\hat R_2(z) = {\cal F}(z)\hat S + k_zr_m,\eqno(7.1)$$
where $\hat S := 1/S$ and $k_z = n_x = [(1-z)/(a+d)]_i$. By interchanging the roles of males and females, the optimal strategy for (5.3.1) can be obtained from the appropriate case in \S 5.2. In particular, with
$$1 = \nu(a+d) + \epsilon\qquad \nu \hbox{ a nonnegative integer}\qquad 0 \leq \epsilon < a+d,\eqno(7.2)$$
and choose $\ell$ so that $\omega := \ell(a+d) < 1-\delta$ but $(\ell+1)(a+d) \geq 1-\delta$, analogous to (6.1--2). Then the optimal return for $\hat R_2$, whence for $R_2$,  must occur at one of
$$0,\ a+d,\ 2(a+d),\ldots,\ell(a+d) = \omega,\ 1.\eqno(7.3)$$
For example, by this procedure one obtains the analogue of (6.a.6) for Model 2 with step-function males and linear females
$$f(z) = \beta(z-\delta);\eqno(7.4)$$
namely, the optimal strategy occurs at the $x$ value $X = 0$, $1-\omega$, or 1 according as:
$$\displaylines{R_2(0) \geq R_2(1)\qquad\&\qquad \beta(a+d) \geq Sr_m,\qquad\hbox{then}\qquad X=0;\cr
R_2(1) > R_2(0)\qquad\&\qquad \beta(a+d) \geq Sr_m,\qquad\hbox{then}\qquad X=1;\cr
\hfill R_2(0) \geq R_2(1)\qquad\&\qquad \beta(a+d) < Sr_m,\qquad\hbox{then}\qquad X=1-\omega;\hfill\llap(5.3.5)\cr
R_2(1) \geq R_2(0)\qquad\&\qquad \beta(a+d) < Sr_m,\qquad\&\qquad\cases{\delta < \epsilon, &then $X = 1-\omega$;\cr
\delta \geq \epsilon,& then $X=1$,\cr}\cr}$$
where $\epsilon$ is as in (7.2).

Due to the asymmetry between males and females in Model 1, however, interchanging the roles of males and females does not convert the current cases into those of \S 6. Since, for the return functions of \S 3, optimal investment always requires equitbale partition of resources amongst several individuals of the same gender, $R_1$ can be obtained by maximizing
$${nr_m \over \mu+n}(\phi+k) + kf\left({1-x\over k}\right),$$
with respect to $n$ and $k$. As $n/(\mu+n)$ is an increasing function of $n$, one gets
$${n_xr_m \over \mu+n_x}(\phi+k) + kf\left({1-x\over k}\right),\eqno(7.6)$$
where $n_x = [x/(a+d)]_i$ and
$$R_1(x) = {n_xr_m \over \mu+n_x}(\phi+k_{1-x}) + k_{1-x}f\left({1-x\over k_{1-x}}\right),\eqno(7.7)$$
where $k_{1-x}$ optimizes (7.6). Now, $C_x := n_xr_m/(\mu+n_x)$ is constant on subintervals of the form
$$[0,a+d),\ \bigl[(a+d),2(a+d)\bigr),\ldots,\bigl[\ell(a+d),1-\delta\bigr)\eqno(7.8)$$
where $\ell$ is as above. So, if $x < y$ within one of these subintervals, then
$$\eqalign{R_1(x) &= C_x(\phi+k_{1-x}) + k_{1-x}f\left({1-x\over k_{1-x}}\right)\cr
&\geq C_x(\phi+k_{1-y}) + k_{1-y}f\left({1-x\over k_{1-y}}\right)\qquad\hbox{by the optimality of $k_{1-x}$}\cr
&\geq C_y(\phi+k_{1-y}) + k_{1-y}f\left({1-y \over k_{1-y}}\right)\qquad\hbox{since $f$ is an increasing function}\cr
&= R_1(y),\cr}$$
i.e., $R_1$ is nonincreasing on each subinterval of (7.8), while on $[1-\delta,1]$ $R_1$ takes its maximal value at 1. Hence, the optimal return is achieved at one of $x$ values of (7.3) after all.

Thus, for step-function males, the optimal return for both Models 1 \& 2 must occur at one of the $x$ values listed in (7.3) and to this extent the situation for both models is analogous to \S 6. While the analogy is complete for Model 2, to underscore the difference for Model 1 between \S\S 6 and 7, recall that for linear males and step-function females, the optimal strategy involved investment in at most one male. Consider now Model 1 for step-function males and linear females. With notation as above, on $[1-\delta,1]$, $R_1(x) = C_x\phi$ while for $z > \delta$ (i.e., $x < 1-\delta$) one must optimize
$$C_x(\phi+k) + \beta(z-k\delta) = C_x\phi + (C_x - \beta\delta)k + \beta z,$$
with respect to $k$, subject to $k < z/\delta$. On $[0,a+d)$, $C_x = 0$ and the coefficient of $k$ is negative. Whenever the coefficient of $k$ is negative, the optimal value of $k$ is 1. But when the coefficient is positive, the optimal value of $k$ is the largest possible value subject to the constraint $k < z/\delta$. The coefficient of $k$ is positive iff $C_x > \beta\delta$, i.e., iff the net return from the investment in males per female in the post-release population exceeds the total cost due to female packaging cost per translocated female. If so, then it pays to invest in more than one linear female, whence the case of step-function males and linear females differs in this respect from that of linear males and step-function females.

Optimization of (7.6) with DMR or Sigmoid females to obtain $R_1$ will be considered in the more general setting of \S 9.
\vskip 24pt
\noindent {\section 8. LINEAR MALES AND FEMALES}
\vskip 12pt
Suppose that the individual return functions for males and females are both linear:
$$m(x) = \cases{0,& if $x \in [0,d]$;\cr a(x-d),& if $x \in [d,1]$;\cr} \hskip 1.25in
f(z) = \cases{0,& if $z \in [0,\delta]$\cr \alpha(z-\delta),& if $z \in [\delta,1]$.\cr}\eqno(8.1)$$
From (3.b.3), ${\cal M}(x) = m(x)$ and ${\cal F}(z) = f(z)$.
If $d+\delta \geq 1$, then one can only invest in either males or females. The returns for Model 1 will be $a(1-d)\phi/(\mu+1)$ and $\alpha(1-\delta)$ respectively.

Now suppose $d+\delta < 1$, i.e., $d < 1-\delta$. Consider first Model 2. Then (4.2) is
$$R_2(x)=\cases{\alpha(1-x-\delta),&if $x \in [0,d]$;\cr
 \psi(x),&if $x \in [d,1-\delta]$;\cr
a(x-d)S,&if $x \in [1-\delta,1]$;\cr}\eqno(8.2)$$
where
$$\psi := a(x-d)S + \alpha(1-x-\delta) = (aS - \alpha)x + \alpha(1-\delta) - adS.\eqno(8.3)$$
Note that $R_2$ is continuous and piecewise linear. $R_2$ decreases linearly on $[0,d]$, increase linearly on $[1-\delta,1]$, and either increases or decreases linearly (as $\psi(x)$) on $[d,1-\delta]$. Thus, there are no local maxima on $(0,1)$ and the maximum of $R_2$ occurs at $x=0$ or 1. Hence, if 
$$R_2(1) = a(1-d)S > \alpha(1-\delta) = R_2(0),\hbox{ equivalently } {aS \over \alpha} > {1-\delta \over 1-d},\eqno(8.4a)$$
choose $x=1$, i.e., assign all resources to a single male; while if
$$R_2(1) = a(1-d)S < \alpha(1-\delta) = R_2(0),\hbox{ equivalently } {aS \over \alpha} < {1-\delta \over 1-d},\eqno(8.4b)$$
choose $x=0$, i.e., assign all resources to a single female. Thus, a very simple strategy results.

Now consider Model 1. On $[0,d]$, $M(x;n;x_1,\ldots,x_n) = 0$ and when investing purely in females the optimal strategy is to invest in a single individual, so $R_1(x) = R_2(x) = R_3(x)$ on $[0,d]$. On $[1-\delta,1]$, $F(z;k;z_1,\ldots,z_k) = 0$, so (4.1) reduces to $M(x;n;x_1,\ldots,x_n)S_P = a(x-nd)[\phi/(\mu+n)]$, $1 \leq n \leq x/d$. This expression is maximized for $n=1$, so $R_1(x) = a(x-d)\phi/(\mu+1) = [\mu/(\mu+1)]R_2(x) = R_3(x)$ on $[1-\delta,1]$. 

On $(d,1-\delta)$, one must maximize the expression
$$a(x-nd){\phi+k \over \mu+n} + \alpha(1-x-k\delta)\qquad 0 < n < {x \over d},\qquad 0  k < {1-x \over \delta},$$
noting that both $n=0$ and $k=0$ are non optimal. The first summand is maximized with respect to $n$ by choosing $n=1$, yielding
$$a(x-d){\phi+k \over \mu+1} + \alpha(1-x-k\delta),\qquad 1 \leq k < {1-x \over \delta},\eqno(8.5)$$
which is linear in $k$. The coefficient of $k$ is 
$${a(x-d) \over (\mu+1)} - \alpha\delta,\eqno(8.6)$$
which is an increasing linear function of $x$ on $(d,1-\delta)$, initially negative and with zero at
$$x^* = {\alpha \over a}\delta(\mu+1) + d > d.\eqno(8.7)$$
When (8.6) is negative, (8.5) is maximized by choosing $k=1$. But when (8.6) is positive, (8.5) is maximized by choosing the largest value of $k$ subject to $k < (1-x)/\delta$, , i.e., the largest $k$ satisfying $x < 1- k\delta$; so if $ x \in \bigl[1-(\ell+1)\delta,1-\ell\delta\bigr)$, choose $k=\ell$.

For simplicity, first suppose that $x^* \geq 1-\delta$, i.e.,
$${\alpha \over a} \geq {1-d-\delta \over \delta(\mu+1)},\eqno(8.8)$$
in which case (8.5) is maximized by choosing $k=1$ at each $x \in (d,1-\delta)$ and Model 1 reduces to Model 3. Substituting $(\phi+1)/(\mu+1)$ for $S$ in (8.3), the result, denoted $\psi_3$, is still linear in $x$. We have already noted that $R_3 = R_2$ on $[0,d]$ and that $R_3 = [\mu/(\mu+1)]R_2(x)$ on $[1-\delta,1]$. Hence, $R_3$ is, like $R_2$, piecewise linear on $[0,1]$ and continuous at $x=d$. But unlike $R_2$, $R_3$ is discontinuous at $x=1-\delta$, due to the discontinuous behaviour of $S_P$ there. When $x < 1-\delta$, one invests $1-x$ in a single female and this female contributes to $S_P$. The limiting return for $R_3$ as $x$ approaches $1-\delta$ from the left is $\lim_{x\to(1-\delta)^-}\,R_3(x) = \lim_{x\to(1-\delta)^-}\,\psi_3(x) = a(1-\delta-d)(\phi+1)/(\mu+1) > a(1-\delta-d)\phi/(\mu+1) = R_3(1-\delta)$, i.e., the actual return at $1-\delta$ is less because one can no longer invest in the female and it drops out of the factor $S_P$. By the same argument as for Model 2, one deduces that there is no local maximum for $R_3$ on $(0,1)$. Thus, the optimal return for $R_3=R_1$, granted (8.8), occurs at $x=0$ or 1; unless $\psi_3$ is increasing, in which case it is possible that the limiting strategy $\lim_{x\to(1-\delta)^-}\,R_3(x)$ exceeds both $x=0$ and 1. If so, there is a strategy that assigns $1-\delta-\epsilon$ to the male and $\epsilon+\delta$ to the female which is more optimal than either $x=0$ or $x=1$. Decreasing $\epsilon$ increases the overall return as long as $\epsilon > 0$. In this case, there would be no actual optimal strategy ($R_3$ has no maximum value on $[0,1]$), though one can achieve a return arbitrarily close to the limiting strategy (which is the supremum, the least upper bound, of $R_3$ on $[0,1]$). This possibility arises only if $\psi_3$ is increasing, i.e., $a/\alpha > (\mu+1)/(\phi+1)$, which is not at all implausible, and if the limiting strategy exceeds the strategies at $x=0$ and 1. The limiting strategy will exceed $R_1(1)$ only when $\delta < (1-d)/(1+\phi)$, which will fail when the number of females in the target population is sufficiently large.

Returning now to the general scenario for Model 1, $R_1(x) = R_3(x)$ on $[0,x^*)$ with a single female invested in on this subinterval. At $x=x^*$, (8.6) is zero and $k$ drops out of (8.5) and $R_1(x^*)$ is independent of the value one chooses for $k$. Suppose $x^* \in \bigl[1-(k^*+1)\delta,1-k^*\delta\bigr)$. Specify $k=k^*$ at $x^*$. 

For $x \in (x^*,1-\delta)$, (8.6) is positive whence (8.5) is maximized by choosing $k$ as large as possible subject to $k < (1-x)/\delta$. Denote this value of $k$ by $k_{1-x}$. For linear returns, one may take the partition of resources $1-x$ amongst $k_{1-x}$ individuals as equitable partitioning. Observe that on $I_k := \bigl[1-(k+1)\delta,1-k\delta\bigr) \cap (x^*,1-\delta)$, $k_{1-x} = k$ and $R_1$ is linear in $x$:
$$R_1(x) = a(x-d){\phi+k_{1-x} \over \mu+1} + \alpha(1-x-k_{1-x}\delta).\eqno(8.9)$$
Thus, $R_1$ is: piecewise linear on $[0,1]$; continuous on $[0,x^*)$; but, if $k^* > 1$, discontinuous at $x = x^*$ , $1-k^*\delta,\ 1-(k^*-1)\delta,\ldots,1-\delta$. Note that the coefficient of $x$ in (8.9) depends on $k_{1-x}$ and decreases with increasing $x$. If, therefore, this coefficient is negative at $x^*$, then it is negative throughout $[x^*,1-\delta)$ and $R_1$ is piecewise linear and decreasing on this subinterval. It also follows that $R_1$ is decreasing linear on $(d,x^*]$ (since $R_1$ is obtained on this subinterval by substituting 1 for $k_{1-x}$ in (8.9)). In this case, the optimal return occurs at $x=0$ or 1. If, on the other hand, the coefficient of $x$ in (8.9) is positive on $[x^*,1-k^*\delta)$, then $R_1$ is linear increasing on this subinterval and either increasing or decreasing on $\left[1-k^*\delta,1-(k^*-1)\delta\right)$, and so on. Once $R_1$ becomes decreasing on one of these subintervals, it is decreasing on the remaining such subintervals. In this case, the optimal return can only occur at 
$$x=0,\ x^*,\ 1-k^*\delta,\ 1-(k^*-1)\delta,\ldots,1-2\delta,\ 1\ (R_1(1) > R_1(1-\delta)).\eqno(8.10)$$
But as we saw in the case when $x^* > 1-\delta$, it may be that $R_1$ does not, because of its discontinuities, have a maximum value on $[0,1]$, but rather approaches a supremum, achieved as 
$$\lim_{x \to y^-}R_1(x),\qquad\hbox{for } y = 1-k^*\delta,\ 1-(k^*-1)\delta,\ldots,1-\delta.\eqno(8.11)$$
In this case, one must examine the returns at $x=0$, 1 and at any other point of (8.10) for which $R_1$ is decreasing on the subinterval that point starts, and the limiting strategy at any point of (8.11) for which $R_1$ is increasing on the subinterval that point terminates. The limit of $R_1(x)$ as $x$ approaches $1-k\delta$ is just $a(1-k\delta-d)(\phi+k)/(\mu+1)$, which exceeds $R_1(1)$ if and only if $\delta < (1-d)/(\phi+k)$.

With $d=0.05$, $a=5$, $\delta=0.05$, $\alpha=1$, $\phi=10$, $\mu=5$, one has $R_1(1) = 7.92$, $R_1(0) = 0.95$, $x^* = 0.11$, $R_1$ is increasing on each of the relevant subintervals $I_k$ and so one must examine the limit of $R_1$ as $x$ approaches the righthand endpoint from the left. One finds that as $x$ approaches each of $0.8$ and $0.85$ from the left, the limit of $R_1(x)$ is 8.75, this value being the supremum of $R_1$ on $[0,1]$ (the other limiting values are strictly less than 8.75). In this case, there is no optimal strategy, only an optimal limiting strategy: take $x$ close to but less than 0.8 (with $k=4$) or close to but less than 0.85 (with $k=3$). In practice one can choose an $x$ close to either 0.8 or 0.85 that yields a return near 8.75, which is greater than the return of 7.92 at $x=1$. Whether such differences in return will prove of practical significance is another question.
\vskip 24pt
\noindent {\section 9. DMR/SIGMOID MALES AND FEMALES}
\vskip 12pt
Finally, we suppose that males and females each have either DMR or sigmoid returns (in any combination). We may treat these returns together since by the arguments of \S 3.d--e the optimal strategy for investing given resources in a gender with either kind of return is equitbale division of those resources amongst the number of individuals characterized as in (3.d.7--8). The packaging costs are $d$ for males and $\delta$ for females and we suppose that $d+\delta < 1$ so that investment in both genders is possible. The male individual return function $m$ is either DMR or sigmoid and therefore dtermines a number $t$ satisfying (3.d.7). Similarly, the individual female return function $f$ is either DMR or sigmoid and also determines a number satisfying (3.d.7) which we denote by $T$, i.e., $T$ is the unique number satisfying $f'(T) = f(T)/T$.

For Model 2, the overall return function $R_2$ is given by (4.2), which can be written
$$R_2(x) = n_xm\left({x \over n_x}\right)S + k_zf\left({z \over k_z}\right),\eqno(9.1)$$
where $z=1-x$, $n_x$ is the optimal number of males in which to invest resources $x$ equitably, $k_z$ is the optimal number of females in which to invest resources $z=1-x$ equitably, and it is understood that the first summand vanishes on $[0,d]$ and the second on $[1-\delta,1]$. We resorted to numerical solution of this model. For any $x$, one can find $n_x$ using (3.d.8) when one knows $t$ and $k_z$ using the analogue of (3.d.8) for $T$. Alternatively, one can compute $nm(x/n)$ for each $n$ satisfying (A.2) and select the optimal $n$ and $kf(z/k)$ for each $k$ satisfying $k\delta < z$ and select the optimal $k$. Repeating this process for a suitable range of $x$ values provides a graph of $R_2$ from which one can determine the $x$ value (to any desired degree of accuracy) yielding the optimal strategy.

But, one can make the following approximations:
$$n_x \approx v_x = x/t \hskip 1.25in k_z \approx z/T,\eqno(9.2)$$
Substituting into (9.1) yields the following approximation to $R_2$:
$$R^a_2(x) = \cases{(1-x)f'(T),& if $x \in [0,d]$;\cr
\bigl(m'(t)S - f'(T)\bigr)x + f'(T)=: A(x),& if $x \in (d,1-\delta)$;\cr
xm'(t)S,& if $x \in [1-\delta,1]$.\cr}\eqno(9.3)$$
Note that extending $A(x)$ to all of $[0,1]$ gives the line joining the two points $\bigl(0,R^a_2(0)\bigr)$ and $\bigl(1,R^a_2(1)\bigr)$. $R^a_2$ is discontinuous at $x=d$ and $1-\delta$ and piecewise linear on $[0,1]$: on $[0,d]$ it decreases linearly from $R^a_2(0) = A(0)$ to $(1-d)f'(T) < A(d)$ at $x=d$; on $(d,1-\delta)$, $R^a_2(x) = A(x)$ either increases or decreases linearly towards $A(1-\delta) > R^a_2(1-\delta)$; on $[1-\delta,1]$, $R^a_2(x)$ increases linearly to $R^a_2(1) = A(1)$. It follows that $R^a_2$ achieves its maximal value at either $x=1$ or $x=0$ according as $A(x)$ is increasing or decreasing, respectively, i.e., $R^a_2$ attains its maximum at
$$x = \cases{1,& if $m'(t)S > f'(T)$;\cr 0,& if $m'(t)S < f'(T)$.\cr}\eqno(9.4)$$
Thus, if the approximation (9.3) is effective, Model 2 has the simple solution (9.4), which is to invest all resources equitably in males if $m'(t)S$ is greater than $f'(T)$, and equitably in females otherwise. Recall from \S 6.b that the approximation (9.2) never underestimates the returns, so $R_2(x) \leq R^a_2(x)$. 

Numerical examples using the function $g$ of (3.d.10) suggest that the rule (9.4) is not without merit for predicting the optimal strategy for Model 2. Further study of the approximation (9.2) may be worthwhile for particular applications. That said, (9.4) is not perfect. With $S= 1.5$, $d = 0.2$ and $m(x) = 6\left(1-(d/x)\right)$, and $\delta = 0.07$ and $f(z) = (1.5)\left(1 - (\delta/z)^3\right)$, i.e., both $m$ and $f$ of the form $g$ in (3.d.10), (9.4) predicts that the optimal strategy occurs at $x=1$. $R_2(1) = 10.8$, but, from the graph of $R_2$, the maximal value is $R_2(0.89) = 11.02$, indicating that a slightly higher return is achieved by investing some (0.11) resources in a single female. In this example, (9.4) fails because (9.2) overestimates the return at $x=1$ ($R^a_2(1) = 11.25$) excessively. Whether such differences in return prove of practical significance is another issue.  

Now consider Model 1. Because resources must be equitably divided amongst individuals of a given sex, see \S 3.d, $R_1$ can be obtained by optimizing the expression
$$nm\left({x \over n}\right)\left({\phi + k \over \mu + n}\right) + kf\left({1-x \over k}\right),\qquad 0 \leq n < {x \over d}\qquad 0 \leq k < {1-x \over \delta},\eqno(9.5)$$
where it is understood that $n=0$ only on $[0,d]$ where the first summand is understood to vanish and $k=0$ only on $[1-\delta,1]$ where it is understood the second summand vanishes. 

The factor $K := \phi + k$ in the first summand of (9.5) merely scales the magnitude of this summand and does not affect the optimization of the first summand with respect to $n$. The first summand in (9.5) is therefore of the same form as in (6.b.6). Replacing (6.b.8) with
$$\psi_x(v) := vm\left({x \over v}\right)\left({K \over \mu + v}\right),$$
the analysis proceeds exactly as in \S 6.b with the result that $\psi_x$ has a unique relative maximum on its domain $(0,x/d)$, which is its actual maximum, occuring at $v_x = x/\tau_x$, where $\tau_x$ solves (6.b.12). Hence, the optimal value $n_x$ of $n$ is the one of the two integers closest to $\tau_x$ yielding the greatest value of the first summand in (5.5.2). Of course, it is always the case that $n_x = 0$ on $[0,d]$ and $n_x = 1$ on $(d,2d]$. After optimization with respect to $n$, (9.5) becomes 
$$b_x(\phi+k) + kf\left({1-x \over k}\right),\eqno(9.6)$$
where
$$b_x := m\left({x \over n_x}\right)\left({n_x \over \mu + n_x}\right),\eqno(9.7)$$

It remains to optimize (9.6) with respect to $k$. On $[0,d]$, $b_x$ is zero and (9.6) reduces to $kf\bigl((1-x)/k\bigr)$, which is optimized as in (3.d.7--8), with $f$ and $T$ taking the places of $m$ and $t$ respectively. For $x \in [1-\delta,1]$, $k=0$ and $R_1(x) = b_x\phi$. On $(d,1-\delta)$, to determine $R_1$ it suffices to optimize
$$b_xk + kf\left({1-x \over k}\right),\eqno(9.8)$$
noting $b_x > 0$. Define 
$$\xi_x(w) := b_xw + wf\left({1-x \over w}\right),\eqno(9.9)$$
on $(0,(1-x)/\delta)$. Then,
$$\displaylines{\hfill\xi'_x(w) = b_x + f\left({1-x \over w}\right) - f'\left({1-x \over w}\right)\left({1-x \over w}\right)\hfill\llap(9.10{\rm a})\cr
\noalign{\vskip 6pt}
\hfill\xi''_x(w) = f''\left({1-x \over w}\right)\left({(1-x)^2 \over w^3}\right).\hfill\llap(9.10{\rm b})\cr}$$
Thus, $\xi_x$ is concave down on its domain, and since $\xi_x$ is assumed smooth, if there is  a relative maximum for $\xi_x$ it is unique and also the maximum of $\xi_x$ on its domain. Suppose then that $\xi_x(w_x) = 0$ and define 
$$\sigma_x := {1-x \over w_x}\eqno(9.11)$$
so that, from (9.10a),
$$b_x = f'(\sigma_x)\sigma_x - f(\sigma_x)\eqno(9.12)$$
is the defining equation for $\sigma_x$. Since $b_x > 0$, one requires $f'(\sigma_x) > f(\sigma_x)/\sigma_x$, whence, from the shape of the graph of $f$, $\sigma_x < T$, where as above $T$ is the solution of (3.d.7) for $f$, i.e.,
$$\delta < \sigma_x < T.\eqno(9.13)$$
Thus, $w_x = (1-x)/\sigma_x > (1-x)/T$, i.e., in Model 1 one expects to invest in more females than one would just to optimize the second summand of $R_1$.

If $f$ is of the form $h = \kappa(x-\delta)^s$ in (3.d.10), (9.12) gives
$$b_x = \kappa(\sigma_x - \delta)^{s-1}\bigl(\sigma_x(s-1) + \delta\bigr),\eqno(9.14)$$
which is not explicitly solvable for $\sigma_x$. On the other hand, if $f$ is of the form $g = \kappa(1 -(\delta/x)^s)$ in (3.d.10), (9.12) yields
$$\sigma_x = \root s \of{s+1}\delta\root s \of{{\kappa \over b_x+\kappa}} = T\root s \of{{\kappa \over b_x+\kappa}} < T.\eqno(9.15)$$
Thus, for functions of the form $g$, one obtains an explicit expression for $\sigma_x$. To determine the value of $k$ maximizing (9.8), one need only test the two integers closest to $\sigma_x$ and choose the one yielding the largest value of (9.8).

If therefore, $m$ is of the form $h$, or $g$ with $s=1$, and $f$ is of the form $g$, one can compute $t$, $T$, $\tau_x$ and $\sigma_x$ explicitly, which reduces the amount of numerical analysis required to determine $R_1$. Whatever the forms of $m$ and $f$, however, for any value of $x$ only finitely many numerical computations are required to compute $R_1(x)$; specifically, for a given value of $x$, compute the first summand of (9.5) for the allowed values of $n$ and thereby determine $n_x$, then compute (9.5), with $n_x$ substituted for $n$, for the allowed values of $k$ and thereby determine $k_x$, the optimal value of $k$, whence $R_1(x)$. 

Numerical studies using the DMR functions of (3.d.10) with various model parameters revealed no obvious restrictions on the value of $x$ yielding the optimal return. To give but one example, using the function $g$ of (3.d.10) for both males and females, with $m(x) = 6(1 - (0.1/x))$ and $f(z) = (1.5)(1-(0.1/z)^3)$, and with $\mu = 5$, the optimal strategy occurs: at $x=1$ when $\phi \geq 19$, with investment in 4 males; at approximately $x=0.87$ when $16 \leq \phi \leq 18$, with investment in 3 males and one female; at values of $x$ in the interval $(0.73,0.77)$ when $9 \leq \phi \leq 15$, with investment in 3 males and two females; at values of $x$ in the interval $(0.44,0.50)$ when $3 \leq \phi \leq 8$, with investment in two males and 4 females; and at values of $x$ in the interval $(0.24,0.28)$ when $0 \leq \phi \leq 2$, with investment in one male and 5 females.

I present one example which illustrates features typified by sigmoid returns. Take both male and female returns to be sigmoid. Take $Z=0.01$ in (6.c.3) for both return functions, but $L = 10$ and $p = 0.3$ for $m$ and $L=2$ and $p = 0.25$ for $f$. Also set $d=0.05$ and $\delta=0.1$ (packaging costs for males and females respectively). With $S=1$, model 2 has its optimal strategy at $x=1$ with investment in 3 males. As $x$ increases from zero to one, $n_x$ increases monotonically from zero to 3, while $k_x$ decreases monotonically from 3 to zero. In Model 1 (with $\phi=\mu=10$ and the same model parameter values), there is no optimal strategy but rather a limiting optimal strategy as $x$ approaches $0.8$ from the left of investing in 2 males and 2 females. Moreover, as $x$ increases from zero to one, $n_x$ increases monotonically from zero to two but $k_x$ no longer decreases monotonically. Once investment in a male exceeds its $p$ value, the return on that male apparently affects the optimal number of females through $S_P$ in this interesting fashion.
\vskip 24pt
\noindent {\section 11. DISCUSSION}
\vskip 12pt
In \S\S 5--10, I have explored at some length various instances of the simple model I have proposed in \S\S 2--4. This mathematical study provides solutions of these scenarios, sometimes in the form of a mathematical result and sometimes just as a numerical procedure, thereby demonstrating the model is tractable.

The biological content of the model is primarily to be found in the representation of reproductive return of an individual as a function of investment, for each gender, with the further assumption that $mS$ and $f$, where $S$ is a suitable measure of mating opportunities of males in the target population, are comparable measures. This assumption is taken over from the literature on sex allocation and parental investment theory, e.g., Frank (1987). I have assumed that the return functions are nondecreasing functions, initially zero to allow for packaging costs, and, with the exception of step-function returns, continuous. Under these assumptions, it was found that optimal investment of resources in a given gender was never found to involve unequal division of resources amongst several individuals, so that the optimal strategy always amounted to equitable partition of resources (though for linear and IMR return functions, the optimal strategy is to invest in a single individual). Linklater's (2003) specific proposal of basing a model on the TW effect could be achieved by the appropriate choice of return functions, perhaps sigmoid returns for males and DMR returns for females in the context of unlimited resources and detailed knowledge of the return on investment of the species in question. Limited resources and incomplete knowledge of returns may entail that other return functions considered in this paper prove practical for modelling purposes.

The return functions studied here are mathematical idealizations of likely measures of reproductive performance of real organisms. For mammals, say, the return functions might represent the number of offspring of an individual of a specific gender in an idealized test population which reach sexual maturity. How well continuous return functions serve to represent discrete returns might depend on the biology of the species in question and the response to levels of investment. Perhaps multi-step-function returns will be more appropriate. One has some indication of the outcomes of the model with multi-step-function returns to the extent that they approximate linear, DMR, or sigmoid returns. In general, however, one must resort to numerical analysis to solve the model with multi-step-function returns, beginning with the essentially combinatorial problem of determining the optimal division of given resources amongst a given number of individuals of the same gender. The solution to this problem will depend on the exact nature of the multi-step function and the optimal division of resources need not be equal division between individuals.

Practical applications require knowledge of the return functions, presumably by experimentation. For endangered species, such would require possibly unjustifiable risks. For black rhino, translocations have not been carried out with varying degrees of investment under experimental conditions. Rather, translocations are typically carried out under a regime of fixed investment. It may be appropriate to model such translocations by step-function males and females. Survival and fecundity estimates of post-release animals could be converted into, say, mean number of offspring/per year (where for males this mean must include inverse weighting by its mating opportunities), and this measure employed as $r_m$ for males and $r_f$ for females in the modelling performed in \S 5 for application to future translocations under the same investment regime. Even such a rudimentary model may be difficult to parametrize however; male fecundity for back rhino is unlikely to be well known since paternity of offspring is rarely known.

Note that the future reproductive success of an individual is dependent upon its current age. I do not model this dependency but implicitly assume that the individuals available for translocation are of optimal age for future reproductive returns. In practice, it may or may not be feasible to meet this assumption. If not met in a known way, it may or may not be straightforward to modify the model to allow for the particular situation.

The fact that in some circumstances the total return function $R_1$ of Model 1 does not a have a maximal {\sl value\/} on the interval $[0,1]$ but only a supremum (least upper bound) is disconcerting. The phenomenon arises from the discontinuity in $R_1$ stemming from the numerator of $S_P$. In particular, an arbitrarily small investment in a female beyond the packaging costs makes a contribution to $R_1$ through its contribution to $S_P$. When this investment in that female decreases to an investment at the level of packaging costs, that investment will optimally be redirected to another female and the decrease in the number of females invested in causes $S_P$ to change discontinuously. As first noted in \S 8, it may then happen that the value of $R_1$ at the point of discontinuity $y$ is less than the one-sided limit of $R_1$ as $x$ increases towards the discontinuity $y$, whence $R_1(y-\epsilon) > R_1(y)$ for arbitrarily small positive $\epsilon$. The limiting return $\lim_{x\to y^-}R_1(x)$ may turn out to be the supremum of $R_1$ on $[0,1]$, but does not represent a strategy which can be implemented. The best one can do is to choose the optimal strategy for $x = y - \epsilon$, for some suitable and small $\epsilon$. It remains to be seen whether this phenomena will prove of practical import. Theoretically, however, it perhaps raises the issue of whether it is realistic to suppose that returns on female investment increase continuously from zero when investment exceeds the known packaging costs. Suppose instead, as with the (multi)-step-function return, a further threshold level of investment beyond the known packaging costs is required for a nonzero return. If at $x=y$ some female receives exactly the threshold level of investment, as $x$ increases toward $y$, $R_1$ is expected to vary continuously and therefore now take its value at $x=y$ given by this limit, i.e., $R_1(y) = \lim_{x\to y^-}R_1(x)$. When $x$ exceeds $y$, the female in question no longer yields a return and the optimal strategy for $x > y$ is to direct investment to another female already invested in (if such occurs) and the numerator of $S_P$ decreases, discontinuously, by one at $x=y$. Now $\lim_{x\to y^+} \not= R_1(y)$, reflecting this discontinuity at $x=y$, and one expects this limit to be less than $R_1(y)$. Thus, such threshold values would not eliminate discontinuities from $R_1$ but render them innocuous, as in the cases involving step-function returns. As with multi-step-function returns, for a return function with a threshold value beyond packaging costs for a nonzero return, the argument of Appendix A which prescribes equitable partition of resources in a given gender as optimal will be violated.

Recall that the model of this paper was designed to optimize the returns of the translocated animals in line with Linklater's proposal to optimize genetic flow into the target population. As such, the optimal strategy provided by this model is not necessarily the optimal strategy as regards reproductive performance of the target population, as this latter goal may not coincide with that of maximizing the reproductive performance of the translocated individuals. As indicated in the introduction, meta-population management involves a number of issues, of which I have considered just one. To indicate one aspect of this point, consider a scenario in which male returns are sigmoid, with $Z = 0.01$, $L = 10$, $p = 0.3$, and $d = 0.05$ in (6.c.3), and female returns are the DMR function $f(z) = (1.5)[1-(0.1/z)^3]$ of (3.d.10). For Model 2, with $S=1$, the optimal strategy is $x=1$ with investment in three males. For Model 1 ($S=1$) with $\mu \geq 14$, the optimal strategy is still $x=1$ but with investment in 8 males. As $\mu$ decreases, so does the value of $x$ at which the optimal strategy occurs. When $\mu=1$, the optimal strategy occurs at $x = 0.27$ with investment in two males and 7 females. These results are neither surprising nor objectionable. When the number of females in the target population is small, the predictions of the model may serve the population reproductive performance less well. Consider the extreme case when $\phi=0$, whence $S=0$ (other model parameters as above). For Model 2, the optimal strategy is $x=0$ with investment in 6 females. For Model 1, however, the optimal strategy depends upon the value of $\mu$. For $\mu \geq 62$, Model 1 yields the same result as Model 2, but when $\mu=61$ the optimal strategy occurs at $x=0.1$, with investment in one male and 6 females; some investment has been diverted from the 6 females to a single male because, according to Model 1, that investment will, through a male's greater reproductive potential, increase the overall return from the translocated individuals, despite the presence of the 61 males in the target population. Perhaps more surprising, there are large values of $\mu$, beginning with $\mu=44$, for which, according to Model 1, the optimal strategy favours investment in males at the expense of the number of females translocated, e.g., for $\mu=20$ the optimal strategy occurs at $x = 0.36$ with investment in 3 males and five females for a net return of 10.1 versus a return of 7.1 for the strategy of $x=0$. Unless there are extenuating circumstances regarding the genetic constitution of the males of the target population, in a case when there are few females in the target population and Model 1 predicts an optimal strategy favoring investment in males at the expense of the number of females translocated, other considerations may veto the recommendation of this strategy. As remarked previously, however, it remains to be seen whether such circumstances will prove of practical importance, e.g., differences in return between certain strategies may prove of negligible importance in practice.
\vskip 24pt
\noindent {\section ACKNOWLEDGMENTS}
\vskip 12pt
I thank Wayne Linklater (WLL) for suggesting that I try to model his proposal put forward in Linklater (2003) and for subsequent feedback on my attempt. I thank the National Research Foundation of the Republic of South Africa and Nelson Mandela Metropolitan University (Port Elizabeth, RSA) for awarding WLL and myself International Science Liaison Foreign Fellowships which enabled us to spend time together in South Africa where we could discuss his proposal with other scientists, especially those involved in black rhino conservation. In particular, I thank the Centre for African Conservation Ecology at NMMU and its Director Professor Graham Kerley for hosting our visit and providing facilities where I began drafting this work into a manuscript.
\vskip 24pt
\noindent{\section REFERENCES}
\vskip 12pt
\frenchspacing
\vskip 1pt
\hangindent=20pt \hangafter=1
\noindent Brett, R. 1998 Mortality Factors and Breeding Performance of Translocated Black Rhinos in Kenya: 1984-1995. Pachyderm, 26:69 -- 82.
\vskip 1pt
\hangindent=20pt \hangafter=1
\noindent Cameron, E. Z. 2004 Facultative adjustment of mammalian sex ratios in support of the Trivers-Willard hypothesis: evidence for a mechanism. Proceedings of the Royal Society London B, 271:1723--1728.
\vskip 1pt
\hangindent=20pt \hangafter=1
\noindent Cameron, E. Z. \& Linklater, W. L. 2000 Individual mares bias investment in sons and daughters in relation to their condition. Animal Behaviour, 60:359--367.
\vskip 1pt
\hangindent=20pt \hangafter=1
\noindent Cameron, E. Z. \& Linklater, W. L. 2002 Sex bias in studies of sex bias: the value of daughters to mothers in poor condition. Animal Behaviour, 63:F5--F8.
\vskip 1pt
\hangindent=20pt \hangafter=1
\noindent Cameron, E. Z., Linklater, W. L., Stafford, K. J., \& Veltman, C. J. 1999 Birth sex ratios relate to mare condition at conception in Kaimanawa horses. Behavioral Ecology, 10:472--475
\vskip 1pt
\hangindent=20pt \hangafter=1
\noindent Carranza, J. 2002 What did Trivers and Willard really predict? Animal Behaviour, 63:F1--F3.
\vskip 1pt
\hangindent=20pt \hangafter=1
\noindent Clark, A. B. 1978 Sex ratio and local resource competition. Science, 201:163 -- 165.
\vskip 1pt
\hangindent=20pt \hangafter=1
\noindent Fischer, J. \& Lindenmayer, D. B. 2000 An assessment of the published results of animal relocations. Biological Conservation, 96:1 -- 11.
\vskip 1pt
\hangindent=20pt \hangafter=1
\noindent Frank, S. A. 1987 Individual and Population Sex Allocation Patterns. Theoretical Population Biology, 31:47--74.
\vskip 1pt
\hangindent=20pt \hangafter=1
\noindent Haight, R. G., Ralls, K., \& Starfield, A. M. 2000 Designing Species Translocation Strategies When Population Growth and Future Funding Are Uncertain. Conservation Biology, 14:1298 -- 1307.
\vskip 1pt
\hangindent=20pt \hangafter=1
\noindent Hearne J. W. \& Swart, J. 1991 Optimal translocation strategies for saving the Black Rhino. Ecological Modelling, 59:279 -- 292.
\vskip 1pt
\hangindent=20pt \hangafter=1
\noindent Leimar, O. 1996 Life-history analysis of the Trivers and Willard sex-ratio problem. Behavioral Ecology, 7:316 -- 325.
\vskip 1pt
\hangindent=20pt \hangafter=1
\noindent Linklater, W. L. 2003 A Novel Application of the Trivers-Willard Model to the Problem of Genetic Rescue. Conservation Biology, 17:906--909.
\vskip 1pt
\hangindent=20pt \hangafter=1
\noindent Lubow, B. C. 1996 Optimal Translocation Strategies for Enhancing Stochastic Metapopulation Viability. Ecological Applications, 64:1268 -- 1280.
\vskip 1pt
\hangindent=20pt \hangafter=1
\noindent Maguire, L. A. 1986 Using Decision Analysis to Manage Endangered Species Populations. Journal of Environmental Management, 22:345--360.
\vskip 1pt
\hangindent=20pt \hangafter=1
\noindent Maguire, L. A., Seal, U. S., \& Brussard, P. F. 1987 Managing critically endangered species: the Sumatran rhino as a case study, pp. 141 -- 185 in Soul\'e (1987).
\vskip 1pt
\hangindent=20pt \hangafter=1
\noindent McPhee, M. E. \& Silverman, E. D. 2004 Increased Behavioral Variation and the Calculation of Release Numbers for Reintroduction Programs. Conservation Biology, 18:705 -- 715.
\vskip 1pt
\hangindent=20pt \hangafter=1
\noindent Miller, B., Ralls, K., Reading, R. P., Scott, J. M. \& Estes, J. 1999 Biological and technical considerations of carnivore translocation: a review. Animal Conservation, 2:59 -- 68.
\vskip 1pt
\hangindent=20pt \hangafter=1
\noindent Robert, A., Sarrazin, F. \& Couvet, D. 2004 Releasing Adults versus Young in Reintroductions: Interactions between Demography and Genetics. Conservation Biology, 18:1078 -- 1087.
\vskip 1pt
\hangindent=20pt \hangafter=1
\noindent Sheldon, B. C. \& West, S. A. 2004 Maternal Dominance, Maternal Condition, and Offspring Sex Ratio in Ungulate Mammals. The American Naturalist, 163:40 -- 54.
\vskip 1pt
\hangindent=20pt \hangafter=1
\noindent Soul\'e, M. E. (ed.) 1987 Viable Populations for Conservation. Cambridge University Press, Cambridge.
\vskip 1pt
\hangindent=20pt \hangafter=1
\noindent Tenhumberg, B., Tyre, A. J. Shea, K., \& Possingham, H. P. 2004 Linking Wild and Captive Populations to maximize Species Persistence: Optimal Translocation Strategies. Conservation Biology, 18:1304 -- 1314.
\vskip 1pt
\hangindent=20pt \hangafter=1
\noindent Trivers, R. L. \& Willard, D. 1973 Natural Selection of Parental Ability to Vary the Sex ratio of Offspring. Science, 179:90--92.
\vskip 1pt
\hangindent=20pt \hangafter=1
\noindent Wang, J. 2004 Application of the One-Migrant-per-Generation Rule to Conservation and Management. Conservation Biology, 18:332 -- 343.
\vskip 1pt
\hangindent=20pt \hangafter=1
\noindent Wootton, J. T. \& Bell, D. A. A Metapopulation Model of the Peregrine Falcon in California: Viability and Management Strategies. Ecological Applications, 2:307 -- 321.
\vskip 24pt
\noindent {\section APPENDIX A}
\vskip 12pt
In this appendix, I present a mathematical result useful for obtaining the optimal division of resources $x$ invested in a single gender. I use the notation appropriate for males.

Let $m$ denote an individual return function, with packaging costs $d$. For fixed resources $x$, let ${\cal P}_n(x)$ denote the set of partitions of $x$ into $n$ amounts. A particular partition $\sum_{i=1}^n\,x_i=x$, $x_i \in [0,1]$, may be denoted $P(x;x_1,\ldots,x_n) \in {\cal P}_n(x)$. Write $y_j :=  x_j - d$, $i=1,\ldots,n$. Since it is wasteful to invest no more than $d$ in an individual, one may suppose that $x_j > d$, i.e., $y_j > 0$, and 
$$x = \sum_{i=j}^n\,x_j = \left(\sum_{j=1}^n\,y_j\right) + nd,\eqno({\rm A}.1)$$
whence
$$nd < x,\eqno({\rm A}.2)$$
which constrains the possible values of $n$. Thus, $y_j \in (0,x-nd)$. Put 
$$M(x;n;y_1,\ldots,y_n) := \sum_{j=1}^n\, m(y_j+d).\eqno({\rm A}.3)$$
One can perform the optimization of $M(x;n;x_1,\ldots,x_n)$ with respect to $x_1,\ldots,x_n$ and $n$ as follows.

First, fix $n$ subject to the constraint (A.2). The problem is then to first choose the $y_j$ so as to maximize $M(x;n;y_1,\ldots,y_n)$, and then to optimize over the allowed values of $n$. Assuming $M(x;n;y_1,\ldots,y_n)$, without constraints, is a $C^2$ function of $(y_1,\ldots,y_n)$, each $y_j \in (0,x-nd)$, then we have a standard optimization problem with constraints. Put $h(y_1,\ldots,y_n) = \sum_{j=1}^n\,y_j - (x-nd)$ and $L(y_1,\ldots,y_n) := M(x;n;y_1,\ldots,y_n) - \lambda h(y_1,\ldots,y_n)$, where $\lambda$ is a Lagrange multiplier. Then,
$${\partial L \over \partial y_j} = m'(y_j+d) - \lambda\hskip .5in j=1,\ldots,n \hskip 1in {\partial L \over \partial \lambda} = h.\eqno({\rm A}.4)$$
Setting these equations to zero gives the solutions for the constrained optimization: $\lambda = m'(y_j+d)$, $j=1,\ldots,n$ and $h=0$, i.e., 
$$m'(y_j+d) = m'(y_k+d)\qquad\hbox{for all\ }j,\ k\hskip 1.25in \sum_{j=1}^n\,y_j = x-nd.\eqno({\rm A}.5)$$
The Hessian $H$ of $L$ is the diagonal matrix
$$H = \diag\bigl(m''(y_1+d),\ldots,m''(y_n+d)\bigr),\eqno({\rm A}.6)$$
evaluated at the solution of (A.5), which is nondegenerate provided each of these second derivatives is nonzero. A sufficient condition to guarantee that the solution is a relative maximum is that $m'' < 0$ at the solution of (A.5).

As a function of $(y_1,\ldots,y_n)$, the domain of interest of $M(x;n;y_1,\ldots,y_n)$ is $y_j \in (0,x-nd)$, $j=1,\ldots,n$, with the constraint $h=0$ restricting $M$ to the hyperplane $\sum_{j=1}^n\,y_j = x-nd$. If $D$ denotes the domain of the constrained function, i.e., $D$ is the hyperplane slice of the open domain of $M$, then its closure is compact and at any point of its boundary, $\overline D \setminus D$, some $y_j$ must be zero. As a constrained function of $(y_1,\ldots,y_n)$, for fixed $x$ and $n$, if $M$ has a maximum on its domain $D$, it must occur at a critical point, which are characterized by (A.5). Otherwise, the `maximum ' of $M$ occurs on the boundary of $D$. Now, the constrained function is continuous, so this maximum can be obtained in a continuous fashion. But, at that boundary point, some $y_j=0$, i.e., one is assigning resources $d$ to the $j$'th individual. This tactic is nonoptimal. It is no worse to reassign the resources $d$ to the other $n-1$ individuals, i.e., there is a partition $P(x;x_1,\ldots,x_{n-1}) \in {\cal P}_{n-1}(x)$ that is at least as optimal as any partition in ${\cal P}_n(x)$. Note that the return function $M(x;n;x_1,\ldots,x_n)$, now viewed as a function of $(n,x_1,\ldots,x_n)$, may change discontinuously with the change from $n$ to $n-1$ in the reassignment of the resources $d$ from the abandoned individual.

The observations of the previous paragraph entail that the optimal investment of resources $x$ occurs at a solution of (A.5) for one of the allowed values of $n$ given by (A.2). In practice, one can begin with the largest allowed value of $n$, solve (A.5) for local maxima, and then proceed to smaller values of $n$, stopping with $n=1$ (investing in no individuals is nonoptimal!). The optimal strategy can be identified from this list of possibilities (if there are no critical points, the optimal strategy is to invest in a single individual).
\vskip 24pt
\noindent{\section APPENDIX B}
\vskip 12pt
We have seen that for the continuous return functions of \S 3 the optimal investment of resources $x$ in a specific gender (say males) may always be construed as equitable division of resources amongst some number $n_x$ of individuals satisfying (A.2): ${\cal M}(x) = n_xm(x/n_x)$. Now ${\cal M}(x) = m(x)$ when $x \in [0,2d]$, with $n_x = 0$ on $[0,d]$ and $n_x=1$ on $(d,2d]$. On $\bigl(sd,(s+1)d\bigr]$, it is possible to invest in $s$ males, and the optimal strategy at $x \in \bigl(sd,(s+1)d\bigr)$ is determined by the relative values of $m(x),2m(x/2),\ldots,sm(x/s)$. Whichever of these functions is largest at such an $x$ is also, by continuity, largest near that $x$. Indeed, one may view the construction of ${\cal M}(x)$ as follows. Construct the graphs of each of the functions $f_n(x) := nm(x/n)$, for those positive integers $n$ satisfying $nd < 1$. This is a finite number of continuous graphs. On $[0,2d]$ all but $f_1(x)$ are zero. So ${\cal M}(x)$ will be continued by $f_1(x) = m(x)$ beyond $x=2d$ until one of $f_n(x)$ equals and then exceeds $m(x)$. This process repeats for the remainder of the domain $[0,1]$ of $\cal M$ whence $\cal M$ is seen to be made up of pieces of the graphs of $f_n(x)$, joined continuously together. Thus, ${\cal M}(x)$ is indeed continuous. 

For linear returns $m(x) = a(x-d)$, $m_n(x) = a(x-nd)$ which is parallel to $m$. Thus, ${\cal M}(x) = m(x)$ as deduced in \S 3b. For an IMR return $m$, observe that $m_n'(x) = m'(x/n) < m'(x)$ by definition of IMR returns. Thus, as with linear returns, no $m_n$ can intersect $m$ and again ${\cal M}(x) = m(x)$ as deduced in \S 3c. This same argument, applied to sigmoid returns, establishes that no $m_n$ can intersect $m$ for $x \leq p$, whence ${\cal M}(x) = m(x)$ on $[0,p]$ as argued in \S 3e.

DMR return functions give a particularly simple picture. With $j > i \geq 1$ and $jd < 1$, suppose that $f_j(x) = f_i(x)$, for some point $x_{i,j} \in (jd,1]$. Since $f'_j(x) = m'(x/j) > m'(x/i) = f'_i(x)$, then at the intersection $x_{i,j}$, $f_j$ does indeed cross over the graph of $f_i$. If, for fixed $i$, $x_{i,j} < x_{i,k}$ whenever $k > j$, then 
$${\cal M}(x) = \cases{m(x),& $x \in [0,x_{12}]$;\cr
f_2(x),& $x \in [x_{1,2},x_{2,3}]$;\cr
f_3(x),& $x \in [x_{2,3},x_{3,4}]$;\cr
\vdots&\vdots\cr}$$
For the functions $h$ and $g$ of (3.d.10), and with $r = 1/s > 1$ for $h$, one finds, respectively,
$$x_{i,j} = {ij(j^r-i^r)d \over ij^r-ji^r} \hskip 1in x_{i,j} = {\root s \of {j^{s+1}-i^{s+1} \over j-i}}\,d={\root s \of {j^s + j^{s-1}i + \cdots + ji^{s-1} + i^s}}\,d.\eqno({\rm B}.1)$$
Note that $ij^r-ji^r > 0$ for $r > 1$, and both the functions in (B.1) are increasing functions of $j$ for fixed $i$ when $j \geq i$. Curiously, both functions are proportional to $d$, but this fact is a coincidence (one can construct other DMR functions that do not share this property). Putting $s=1$ in the formula for $g$, one obtains the particularly simple result that $x_{i,j} = (j+i)d$. Since $x_{i+1,i+2} - x_{i+1,i} = 2d$, in this case
$${\cal M} =\cases{m(x),& $x \in [0,3d]$;\cr
2m(x/2),& $x \in [3d,5d]$;\cr
3m(x/3),& $x \in [5d,7d]$;\cr
\vdots&\vdots\cr}\eqno({\rm B}.2)$$
For the sigmoid function of (6.c.1), one does not obtain an explicit formula for $x_{i,j}$.

\bye